\documentclass[11pt]{JHEP3}%JHEP3

\usepackage{cite,slashed}
\usepackage{amsmath,amstext,amssymb}
\usepackage{graphics}
\usepackage{psfrag}
\usepackage{axodraw}

\def\al{\alpha}

\def\dl{\delta}
\def\eps{\epsilon}

\def\om{\omega}

\def\pa{\partial}

\def\bc{\begin{center}}

\def\ec{\end{center}}
\def\be{\begin{eqnarray}}
\def\ee{\end{eqnarray}}

\title{Non-Equilibrium Field Dynamics of an Honest Holographic Superconductor}
\author{
Xin Gao$^{a,d}$, Matthias Kaminski$^{b}$, Hua-Bi Zeng$^{c,e}$, Hai-Qing Zhang$^{d}$ \\ 
$^{a}$ Max-Planck-Institut f\"ur Physik (Werner-Heisenberg-Institut)
  F\"ohringer Ring 6, \\
  80805 M\"unchen, Germany\\
$^{b}$ Department of Physics, University of Washington
Seattle, WA 98195-1560, USA \\
$^{c}$ Department of Physics, Nanjing University
Nanjing 210093, China \\
$^{d}$ 
State Key Laboratory of Theoretical Physics, Institute of Theoretical
Physics, Chinese Academy of Sciences, P.O. Box 2735, Beijing 100190, China\\
$^{e}$School of Mathematics and Physics, Bohai University
JinZhou 121000, China

Email: \email{gaoxin@mpp.mpg.de, mski@uw.edu, zenghbi@gmail.com,hqzhang@itp.ac.cn}
}

\abstract{
Most holographic models of superconducting systems neglect
the effects of dynamical boundary gauge fields during the process of spontaneous
symmetry-breaking. Usually a global symmetry gets broken.
This yields a superfluid, which then is gauged "weakly" afterwards. In this
work we build and probe the dynamics of a holographic model in which a local boundary symmetry is spontaneously broken instead. We compute two-point functions of dynamical non-Abelian gauge fields in the normal and in the broken phase,
and find non-trivial gapless modes. Our AdS$_3$ gravity dual realizes a 
p-wave superconductor in (1+1) dimensions.
The ground state of this model also breaks $(1+1)$-dimensional parity spontaneously, while the Hamiltonian is parity-invariant. We discuss possible implications of our 
results for a wider class of holographic liquids.
}
\preprint{MPP-2012-77}
\begin{document}
\maketitle

%%%%%%%%%%%%%%%%%%%%%%%%%%%%%%%% INTRO
\section{Introduction}

The gauge/gravity correspondence~\cite{Maldacena:1997re},
also known as AdS/CFT correspondence
or "holography", has recently provided the tools to study models of strongly correlated systems exhibiting superconducting phases~\cite{Gubser:2008px, Hartnoll:2008vx, Hartnoll:2008kx}.
All of these constructions have in common that first a superfluid emerges
from a process which spontaneously breaks a global symmetry. Then the resulting
field theory is "weakly gauged" afterwards. In other words there are no dynamical
gauge fields involved in the symmetry-breaking process, which means that these
models neglect effects of order $\mathcal{O}(1/{g_{\text{YM}}}^2)$ due to loop corrections which would involve photons in a real superconductor 
($g_{\text{YM}}$ is the gauge coupling). This may be seen as a
shortcoming when trying to model
"honest" superconductors in which dynamical gauge fields are present
during the symmetry-breaking.
In this work we address this issue.

Our first main goal in this paper is to study spontaneous symmetry-breaking in presence of dynamical
gauge fields. In this sense we are building an "honest" holographic
superconductor. S-wave superconducting ground states in presence of dynamical gauge
fields have been studied in higher dimensions~\cite{Domenech:2010nf,Montull:2012fy,Silva:2011zz,Montull:2011im}.
In the present work we extend these studies in three ways:
i.)~we examine the dynamics of such a setup, most prominently the two-point
functions of dynamical gauge-fields,
ii.)~we consider a vector order parameter, i.e. a p-wave solution, and
iii.)~we work in $(1+1)$ dimensions.
Note that the Mermin-Wagner theorem is evaded in
the large $N$ limit~\cite{Hartnoll:2009sz,Anninos:2010sq}, allowing spontaneous symmetry-breaking in two spacetime dimensions. The normal phase of $(1+1)$-dimensional
fluids in presence of an Abelian field was studied holographically in~\cite{Hung:2009qk},
but using an interpretation distinct from ours.\footnote{We will return to this work for comparison below.}

Our second main goal in this paper is to study the two-point functions of our dual field theory as mentioned in point i.) above. These are in particular interesting since our setup forces us to work in the so-called "alternative quantization". We are exploring this alternative formulation and its low-energy effective field theory, which we call "non-equilibrium field dynamics". 
Gauge fields in $AdS_3$ can in general have Neumann or Dirichlet boundary conditions. But in this paper we restrict ourselves to a special case where only Neumann conditions are possible. We consider a Yang-Mills action without any other terms, in particular we have no Chern-Simons term. In this special case gauge fields in $AdS_3$ can only have Neumann boundary conditions, so the dual field theory is always written in the alternative quantization.
This is discussed in greater detail below towards the end of this introduction. In this context we also note that we can only study the superconductor (dynamical boundary gauge field), not a superfluid (external non-dynamical boundary gauge field) in our setup. Therefore we can not directly compare the two within our setup. But we find that in presence of our dynamical gauge field a spontaneous symmetry breaking occurs in our model. And formally this symmetry breaking in our holographic superconductor is very similar to the symmetry breaking known from higher-dimensional holographic superfluids.

We will focus here on the p-wave solution to a Yang-Mills action in AdS$_3$-space
\begin{equation}\label{eq:action}
S =\int d^3x \sqrt{-g}\mathcal{L} =-\int d^3x \sqrt{-g} g^{\mu\mu'} g^{\nu\nu'} F^a_{\mu\nu} F^a_{\mu'\nu'}\, ,
\end{equation}
where $g$ is the metric of AdS$_3$, and $F$ is an $SU(2)$ field strength. 
Note that we allow no Chern-Simons term here. This corresponds to forcing the chiral anomaly in the dual gauge theory to zero~\cite{Witten:1998qj,Freedman:1998tz}.
An action such as~\eqref{eq:action} arises in a particular string theory setup, see section~\ref{sec:stringAction} for details. It can be
thought of as a small-field expansion of the non-Abelian Dirac-Born-Infeld action of two
coincident probe D3-branes in a background generated by
$N$ D3 branes with $N\to \infty$~\cite{Karch:2002sh}. For earlier discussions of this brane setup
see~\cite{Constable:2002xt,Hung:2009qk,Jensen:2010em}.\footnote{Closely related to this are the
constructions~\cite{Ammon:2008fc, Basu:2008bh, Ammon:2009fe,Ammon:2009xh}. For a review of the relevant methods see~\cite{Kaminski:2010zu}.}
The Yang-Mills-theory with action~\eqref{eq:action} is interesting to us for three main reasons:

First, the expectation is that we introduce gapless modes into an otherwise simple setup. For example, in \cite{Jensen:2010em} it was found that the duals to AdS$_3$ Maxwell-Chern-Simons theory do not show any hydrodynamic behavior, all correlators are fixed by symmetries. Now let us naively think of our bulk Yang-Mills-theory defined by~\eqref{eq:action} as such a Maxwell-Chern-Simons theory but with vanishing Chern-Simons coupling. When we introduce the p-wave condensate, we introduce gapless modes (emerging from the Goldstone modes) into an otherwise empty system. So the hope is to study separately these gapless modes, possibly analytically.

Second, our p-wave solution breaks parity in $(1+1)$ dimensions. That means we have a parity-breaking ground state from a parity-invariant action. And the effective
field theory description of the field theory dual to this setup needs to incorporate parity-violating terms. In contrast to the Chern-Simons setups in AdS$_5$~\cite{Erdmenger:2008rm,Banerjee:2008th,Son:2009tf}, here the parity-violating transport is not related to an anomaly, but to spontaneous symmetry-breaking.

Third, Yang-Mills-theory in AdS$_3$ is different in that it allows only one
particular quantization: the bulk gauge fields $\mathcal{A}(t,x,z)=A(t,x,z)+a(t,x,z)$
(background and fluctuations) correspond to dynamical gauge fields in
the boundary field theory.
This boundary theory also contains currents, but they are
non-dynamical, as they act as sources. In other words, the currents and
gauge fields have switched their roles compared to the ordinary quantization.
This claim is based on gauge-invariance and the presence of leading
logarithmic terms in the boundary expansion of gravity fields. This allows only
Neumann boundary conditions for the bulk gauge fields.
For a careful analysis see~\cite{Marolf:2006nd, Jensen:2010em},
(see also comments in~\cite{Andrade:2011sx}). One has to carefully
repeat the analysis of~\cite{Klebanov:1999tb} but for vector fields instead of
scalar fields. 

For scalars~\cite{Klebanov:1999tb} showed that the 
two possible quantizations are related by a Legendre transformation 
and change of boundary conditions. 
Also the scalar operator two-point function in one
quantization is simply the inverse of the scalar operator two-point function in the
other quantization.\footnote{Note that changing quantizations
amounts to changing the boundary conditions \emph{and} performing
a Legendre transformation.} 
A generalization to vector fields in $(2+1)$-dimensional conformal gauge theories 
(with AdS$_4$ gravity duals) was carried out in~\cite{Witten:2003ya}:
The two quantizations are related by an $SL(2,\mathbb{Z})$ 
transformation in the bulk. On the gravity side, the two bulk theories are 
dual to each other in the sense that $SL(2,\mathbb{Z})$ is an extension of the 
traditional electric-magnetic duality symmetry.
On the field theory side, there is no obvious symmetry 
relation between the two (respectively corresponding) boundary theories.
They are in general physically inequivalent\footnote{
It has been shown that in the conformally-symmetric case there exists a group-theoretic
equivalence between the two different quantizations~\cite{Dobrev:1998md}.}, but related to each other:
The application of $SL(2,\mathbb{Z})$ to 
one boundary theory takes us to the other boundary theory.
So we can compute observables in one quantization if we know the other.
Now in stark contrast to this, we are interested in a non-conformal
theory with gauge fields in 
$(1+1)$ dimensions where only one quantization exists if we do not allow a Chern-Simons term in our action~\eqref{eq:action}. 
So a possible relation of conserved current Green's functions to 
gauge field Green's functions is less clear here, if it exists at all.\footnote{
According to~\cite{Dobrev:1998md}, for any tensor field (including scalars and vectors) in a conformal theory in any dimension the two quantizations are
related by a Weyl reflection, and two-point functions are related to each other. 
But our model at hand breaks conformal symmetry.} 

In our alternative quantization one has to carefully question some of the 
concepts which are by now well 
established in the normal quantized versions of gauge/gravity.
What we do not doubt is that gauge/gravity is a
correspondence between bulk fields and operators on the boundary.
In this generality we are going to use gauge/gravity in this work and consequently
analyze one-point functions and two-point functions
of our operators, by which we mean our dynamical boundary gauge fields,
denoted by $\mathfrak{A}(t,x)$ for background fields, and $\mathfrak{a}(t,x)$
for fluctuations around the background. So in other words
$A(t,x,z)+a(t,x,z)$ in the bulk corresponds to $\mathfrak{A}(t,x,z)+\mathfrak{a}(t,x,z)$
on the boundary. In this context we are going to make two
further assumptions (which are standard in AdS/CFT):
\begin{enumerate}
\item Gauge-invariant bulk field combinations correspond to physical operators in the boundary field theory.
\item Quasinormal modes of gauge-invariant bulk field combinations coincide with
the poles of two-point functions of physical operators in the boundary field theory.
\end{enumerate}

Our setup with dynamical gauge fields and fixed currents also obscures the
standard interpretation in terms of hydrodynamics. Hydrodynamics is a low-energy effective field theory which is always expressed in terms of dynamical conserved currents (organized in constitutive equations and conservation equations, see e.g.~\cite{Kovtun:2012rj} for a review). In contrast to this, we have to consider a theory with dynamical gauge fields in the present work. We call the low-energy effective field theory of our field theory "non-equilibrium field dynamics", or shorter "field dynamics" (in order to distinguish it from hydrodynamics). When doing linear response in our setup, we have only access to two-point-functions of gauge fields (gauge field propagators), but not to current-current correlators.
We note that this formulation is very similar to the language used in thermal
quantum field theory, see {e.g.}~\cite{Kapusta:2006pm}, and when discussing
non-equilibrium thermal dynamics. Therefore one may hope that the
gravity setting with Neumann boundary conditions provides results which compare
more directly with field theory approaches. We stress again
that the main point of our work is not a comparison of superfluids with superconductors. Instead we want to explore the alternative setup. In particular, we explore what we call Ófield dynamicsÓ, i.e. the low-energy limit of our alternative description. This is analogous to ÓhydrodynamicsÓ in the ordinary description.

We start in section \ref{sec:hydro} with a few thoughts on possible
field theoretic descriptions of systems with dynamical gauge fields in two
spacetime dimensions.
In section \ref{sec:setup} we introduce a concrete holographic setup, the
Yang-Mills action in AdS$_3$ and our symmetry-breaking solution with vector hair.
Then, in section \ref{sec:fluctuations} we present our analysis of fluctuations,
explaining in detail how standard methods need to be adjusted to the
situation in AdS$_3$. Finally we close with a discussion of open tasks
in section~\ref{sec:discussion}. Details on gauge-covariant and
gauge-invariant fields are collected in appendix~\ref{sec:covariantFields}.
In appendix~\ref{sec:check} we cross-check some of our results when 
using the ordinary interpretation with gauge fields being the sources.

%%%%%%%%%%%%%%%%%%%%%%%%%%%%%%%% HYDRO
\section{Dynamical Gauge Fields in $(1+1)$ Dimensions}
\label{sec:hydro}
Although we are going to study a simple system on the gravity side,
there will be several complications on the field theory side. 
We know of no suitable formulation of superfluids or superconductors in $(1+1)$
dimensions expressed in a way which would be directly comparable to our setup. Hence, in this section
we will outline our best estimate for such a suitable description.
In order to disentangle the relevant topics, we separate and discuss them here
from the field theory perspective (as opposed to the gravity perspective).
We will be rather schematic and make qualitative observations here,
before we introduce a concrete model in section~\ref{sec:setup}.

\subsection{Thermodynamics}
As mentioned above, we are going to study gauge fields and their dynamics
in this work. 
Finding a version of hydrodynamics with dynamical gauge fields (instead of the usual dynamical currents) turns
out to be a bit tricky (as discussed below). However, our equilibrium
states will have a fairly simple standard interpretation. For later convenience
we here review the canonical and the grand-canonical ensemble.
The grand-canonical ensemble is defined to be in contact with a particle
reservoir. Therefore it has a fixed chemical potential $\mu$, and the
grand-canonical potential density is $\Omega(T,\mu)$.
The canonical ensemble has a fixed charge density $\rho$ and the
relevant potential is the free energy density $\mathcal{F}(T,\rho)$. A Legendre transformation
takes us from one ensemble into the other
\begin{equation}
\Omega = \mathcal{F} - \mu \rho \, .
\end{equation}
So here we have two distinct setups: First, the grand-canonical ensemble
(at fixed chemical potential $\mu$) which we will associate with the "ordinary" setup later.
Second, there is the canonical ensemble (at fixed charge density $\rho$),
which we will associate with the "alternative" setup later.
In this context these two setups have been used many times before,
and usually one does not talk about this as "two distinct quantizations". But the
two setups are indeed related by a Legendre transform on the field theory side
(plus appropriate choice of boundary conditions fixing either $\mu$ or $\rho$).
Also the gravity duals to both setups are related by a bulk Legendre
transformation (plus a change of boundary conditions), which takes
us from one quantization of the gauge field to the other. This has been first
discussed for a scalar in \cite{Klebanov:1999tb}.
In the stationary state which captures thermodynamics, it is thus
clear how to interpret the two distinct setups, the "ordinary" and the
"alternative" one.

\subsection{Hydrodynamics vs. Field Dynamics}
\label{sec:altNormalHydro}

Let us consider the linear response of a normal fluid at low momentum and frequency
in $1+d$ spacetime dimensions
\begin{equation}
\langle \mathfrak{j}\rangle = G^R_{\mathfrak{j}\mathfrak{j}} \cdot \mathfrak{a} \, ,
\end{equation}
where the angular brackets indicate that $\langle \mathfrak{j}\rangle$ is a one-point function
of the conserved current $\mathfrak{j}=(\mathfrak{j}^t, \mathfrak{j}^\alpha)^T$,
and $\mathfrak{a} = (\mathfrak{a}_t, \mathfrak{a}_\alpha)^T$ is the gauge field, which acts as
a source here. Greek indices run over the $d$ spatial dimensions.
By $G^R_{\mathfrak{j}\mathfrak{j}}$ we mean the $(1+d)\times (1+d)$ matrix of two-point functions of the current $\mathfrak{j}$ with itself. We define this to be the "ordinary" setup. This is the setup
which has been realized in most holographic models (including those of holographic superconductors) to date.
The low-frequency and low-momementum limit of this setup is
simply ordinary "hydrodynamics".

Alternatively, we can define a different system by switching the roles of sources
and one-point functions, so that we get
\begin{equation}\label{eq:<a>}
\langle \mathfrak{a}\rangle = G^R_{\mathfrak{a}\mathfrak{a}} \cdot \mathfrak{j} \, ,
\end{equation}
where now the angular brackets indicate that $\langle \mathfrak{a}\rangle$ is a one-point function
of the gauge field $\mathfrak{a}$, and $\mathfrak{j}$ is a conserved current, which now acts as
source. By $G^R_{\mathfrak{aa}}$ we mean the $(1+d)\times (1+d)$ matrix of two-point functions
of the gauge
field 
$\mathfrak{a}$
with itself. We will refer to this system as the "alternative" setup.
This is the setup which our gravitational setup (with Neumann boundary conditions
on the gravity fields) will correspond to through
the gauge/gravity correspondence. We refer to the low-frequency, low-momentum limit of this setup as
"field dynamics" (in order to distinguish it from the hydrodynamic description above).

It is not obvious, how the ordinary setup is related to the alternative one, i.e. how "hydrodynamics"
is related to "field dynamics".
It is also not \emph{a priori} obvious what physics is captured in the alternative
setup at low frequencies and momenta (i.e. in "field dynamics").
So one of our goals will be to study the low-frequency, low-momentum
fluctuations of our gravity model in the alternative setup. As noted before, in our $AdS_3$ gravity model
we have no access to the ordinary setup. Therefore a direct comparison between the (well-known) ordinary "hydrodynamics" and the (less conventional) alternative "field dynamics" setup is not possible in our case.

Note also that it is not possible to naively invert ordinary hydrodynamics in order
to get an alternative version of hydrodynamics in terms of gauge fields.
In order to see this, let us consider a simple ordinary hydrodynamics
setup with a constitutive equation for the dynamical current
\begin{equation}\label{eq:<j>}
\langle \mathfrak{j}^\mu\rangle = \rho u^\mu + \sigma \left(e^\mu-\nabla^\mu (\frac{\mu}{T})\right)\, ,
\end{equation}
where $e_\mu=u^\nu(\partial_\mu\mathfrak{a}_\nu-\partial_\nu\mathfrak{a}_\mu)$ is the electric field, $\mu$ is the chemical potential which
couples to the charge density $\rho$, and $u^\mu=(1,0)$ is the 2-velocity
(all to first order in derivatives). Let us assume small field fluctuations
$\mathfrak{a}^t$ and
$\mathfrak{a}^x$, to which the system responds by charge fluctuations in $\rho$.
We keep the temperature $T$ and the 2-velocity fixed. Now our goal would
be to solve this equation for $\mathfrak{a}^\mu$ (such that we could write
$\mathfrak{a}^\mu = \dots$,
where $\dots$ are terms which do not explicitly depend on $\mathfrak{a}^\mu$).
Using the conservation equation
$\nabla_\mu \langle \mathfrak{j}^\mu\rangle = 0$, and Fourier-transforming by
$\mathfrak{a}^\mu\to \mathfrak{a}^\mu \exp(-i\omega t+ikx)$, we may rewrite
\eqref{eq:<j>} as a matrix equation
\begin{equation}
\left(
\begin{array}{c}
\langle \mathfrak{j}^t\rangle \\
\langle \mathfrak{j}^x\rangle
\end{array}\right)=
-\frac{i\sigma}{\omega+i D k^2}
\left(\begin{array}{c c}
k^2 & \omega k\\
\omega k & \omega^2
\end{array} \right )
\left(
\begin{array}{c}
\mathfrak{a}^t\\
\mathfrak{a}^x
\end{array}\right)
= {\cal W}\cdot \mathfrak{a}
\, ,
\end{equation}
with the diffusion constant $D = \sigma/\chi$, where $\chi$ is the electric
susceptibility.
For this system to be invertible, we would need the matrix $\cal W$
to have an inverse. But as we easily see, this is not the case since
$\det{\cal W} =0$.

\subsection{Superfluid Field Dynamics}
In ordinary superfluid hydrodynamics we have to add
the phase of the order parameter as a new degree of freedom
compared to the normal-fluid hydrodynamics.
This is due to the Goldstone theorem which tells us that for every spontanously
broken continuous symmetry a massless bosonic field appears in the
spectrum of our theory. Superfluids just below the critical temperature
can be thought of as a mixture of two phases, a superfluid, and a normal-fluid one. 
Once the symmetry-breaking has occured, superconductors can be described as charged superfluids for the purpose of studying low-energy excitations
and transport phenomena. Therefore in the literature superconductors and superfluids are often described
in the formalism of superfluid hydrodynamics. This is the ordinary setup which we will have no access to in
our $AdS_3$ model. Hence we are not able to formulate our results in this common way.

In our alternative setup we still expect new degrees of freedom
since we are still breaking a continuous symmetry spontaneously.
To be more precise, our gravity solution will develop vector hair below a critical temperature,
i.e. we have a non-trivial background gauge field $A_x^1(z)$. This
breaks a residual $U(1)$ gauge symmetry spontaneously. Thus
we expect to see the Goldstone-modes from this symmetry-breaking
in the two-point functions of our dynamical gauge fields.
But it is not clear what the behavior of these modes is going to be.
For example one could study their dispersion relations $\omega(k)$. 
In the broken phase however, we restrict ourselves to the case of vanishing spatial  momentum, 
and compute selected Green's functions. Only
for the normal phase we consider non-zero momentum and find a linear dispersion, as discussed below.

\subsection{Parity-Violating Field Dynamics}
Our gravity solution leads to spontaneous parity-breaking in the dual $(1+1)$ dimensional
field theory. Therefore we also expect to see effects of parity-violation
in our low-frequency, low-momentum modes. Note that parity-violating
hydrodynamics (in the ordinary formulation) for superfluids with a scalar order parameter has been studied in
$3+1$ dimensions~\cite{Bhattacharya:2011tra}. There, parity-breaking leads
to a large number of additional transport coefficients. In principle it should be possible
to carry out a similar analysis in the alternative setup in order to derive a formulation of parity-violating field dynamics. For example one
could use the fluid/gravity framework~\cite{Bhattacharyya:2008jc} and
compute the constitutive relations (presumably for dynamical gauge fields)
and corresponding coefficients explicitly. It is not clear what
to expect from such an analysis. We note that adding a Chern-Simons term to
our action~\eqref{eq:action} would lead to a chiral anomaly on the field theory
side. This would break parity as well and such setups have been examined
recently in~\cite{Jensen:2010em,Zayas:2011dw,Andrade:2011sx}.
However, this would also change the boundary behavior of the gauge fields.

\subsection{Massive Gauge Bosons and the Schwinger model}\label{sec:schwinger}
One may speculate that the physics of dynamical gauge bosons should be
best captured by thermal quantum field theory,
at least at weak coupling.
In this case the correlators
$G^R_{\mathfrak{aa}}$ between gauge bosons $\mathfrak{a}$
discussed around \eqref{eq:<a>}
should be interpreted as gauge boson propagators. Due to the finite temperature
we naively expect our gauge bosons to have a thermal mass. Furthermore, they
should acquire a mass through a Higgs mechanism due to the spontaneous symmetry-breaking. The latter mass contribution should depend on the condensate which breaks the symmetry. So in particular this mass contribution should vanish as we approach
the phase transition from the broken phase. But we will see below that
neither the thermal mass nor the mass from symmetry-breaking can appear
in our setup.

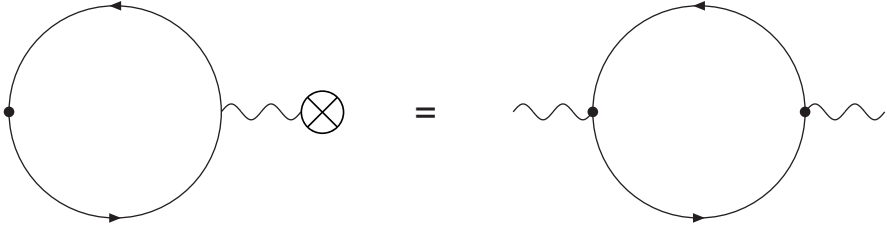
\begin{figure}
\begin{center} \begin{picture}(350,100)(0,0)
\ArrowArc(40,50)(40,0,180)
\ArrowArc(40,50)(40,180,360)
\Vertex(0,50){2}
\Photon(80,50)(110,50){3}{2} 
\SetPFont{Helvetica}{14}
\Text(118.5,50)[]{\Huge $\otimes$}
\PText(157,53)(0)[]{=}
\Vertex(300,50){2}
\Photon(190,50)(220,50){3}{2}
\ArrowArc(260,50)(40,0,180)
\ArrowArc(260,50)(40,180,360)
\Vertex(220,50){2}
\Photon(300,50)(330,50){3}{2}
\end{picture} \end{center}
\caption{\label{fig:PiIsAnomaly}
In $(1+1)$ dimensions the anomaly diagram on the left
is identical to the self-energy diagram for the gauge field on the right.
}
\end{figure}

Already at zero temperature gauge bosons in $(1+1)$ dimensions
are different from higher-dimensional cases. Schwinger discussed quantum electrodynamics in two spacetime dimensions~\cite{Schwinger:1962tp}.
Among other things he found that photons in $(1+1)$ dimensions
can have a mass term which does not break gauge invariance. Furthermore
the self-energy diagram of gauge fields in $(1+1)$ dimensions is identical
to the diagram for the chiral anomaly in $(1+1)$ dimensions, see
figure~\ref{fig:PiIsAnomaly}. The mathematical reason
for this is that there are two $\gamma$-matrices in $(1+1)$ dimensions,
$\gamma^0,\, \gamma^1$, plus the $\gamma^5=\gamma^0\gamma^1$,
and they obey the relation $\gamma^\mu\gamma^5 = -\epsilon^{\mu\nu}\gamma_\nu$,
with the totally antisymmetric symbol in two spacetime dimensions $\epsilon^{\mu\nu}$.
Thus in the fermion loop in figure~\ref{fig:PiIsAnomaly} on the left we can replace the coupling to an external axial current
$i \gamma^\mu\gamma_5$
by the coupling to an external gauge field $-i\gamma_\nu$.
This implies a strong constraint:
If a gauge field in $(1+1)$ dimensions has a mass, then the theory has
a chiral anomaly, and vise versa.

We will use this result below in order to interpret the $(1+1)$-dimensional field theory
which is dual to our AdS$_3$ setup. Note that for a massive
photon in $(1+1)$ dimensions the Ward identity requires a pole in the gauge boson
propagator at two-momentum $k^2=0$~\cite{Schwinger:1962tp}.

%%%%%%%%%%%%%%%%%%%%%%%%%%%%%%%% SETUP
\section{Holographic Setup and Symmetry-Breaking Solution}
\label{sec:setup}
While the previous section was very speculative, in the present 
section we are going to work with a concrete gravity model.
On the dual field theory side, this will realize a p-wave superconductor
in $(1+1)$ dimensions with dynamical electromagnetic fields, i.e.
the alternative setup described above.

%-------------------------------------------------
\subsection{Dirac Born Infeld Action for Coincident D3 Probe Branes}
\label{sec:stringAction}

In the context of probe brane constructions~\cite{Karch:2002sh} --apart from the DBI-action-- the effective brane world volume action also contains a Wess-Zumino term. This leads to the Chern-Simons coupling of the gauge field. However, for particular brane intersections such a Chern-Simons term can be absent and only the Maxwell term remains. We consider a defect brane configuration which realizes this very situation.
\begin{table}[ht]
  \centering
  \begin{tabular}{c|c|c|c|c|c|c|c|c|c|c}
     & t & x & y & z & 4 & 5 & 6 & 7 & 8 & 9  \\
    \hline 
    $D3$ & X & X & X & X & & & & & \\
    \hline
    $D3'$ & X & X & & & X & X & & & \\
  \end{tabular}
  \caption{\label{tab:D3D3'} $D3-D3'$ brane configuration which provides only a Maxwell but no Chern-Simons term. We consider $N\to \infty$ $D3$ branes, and $N_f=2$ $D3'$ probe branes.}
  \label{braneconfiguration}
\end{table}  
This is the $D3-D3'$ defect configuration illustrated in Table \ref{tab:D3D3'}. The metric generated by the $N$ background $D3$ branes is
the standard $AdS_5 \times S_5$ (black brane) metric
\begin{equation}
 ds^2_{D3}=H^{-1/2}(-f(r)dt^2+dx^2+dy^2+dz^2)+H^{1/2}(\frac{dr^2}{f(r)}+r^2d\Omega^2_5) \, ,
\end{equation}
where $f(r)=1-\frac{r_0^4}{r^4}$ , $H=(L/4)^4$ and $
d\Omega_5^2=d\theta^2+\cos^2\theta d\xi^2+\sin ^2\theta dS_3^2$.
Transforming the radial coordinate to $u=r_0/r$, the metric becomes
\begin{equation}
 ds^2_{D3}=\left(\frac{r_0}{L}\right)^2\frac{-f(u)dt^2+dx^2+dy^2+dz^2}{u^2}+\frac{L^2 du^2}{u^2 f(u)}+L^2d\Omega^2_5 \, .
\end{equation}
We consider $N_f$ coincident probe $D3'$ branes with a trivial embedding, i.e. $\theta(r)= 0$. Then the metric induced on the world volume of the $D3'$ branes is
\begin{equation}
 ds^2_{D3'}=\left(\frac{r_0}{L}\right)^2\frac{-f(u)dt^2+dx^2}{u^2}+\frac{L^2 du^2}{u^2 f(u)}+L^2 d\xi^2 \, .
\end{equation}
From this metric we can see that the probe $D3'$ branes cover $AdS_3$ as well as
an $S_1$ cycle inside the five-sphere. The existence of such $AdS_3\times S^1$ embeddings for the $D3'$ branes has been demonstrated in \cite{Constable:2002xt}. 
Furthermore we switch on a $U(N_f)$ gauge field living on our $(1 + 1)$-dimensional defect. In fact our gauge field lives on the world volume of the $D3'$ branes. These probe branes do not wrap any cycle with Ramond-Ramond flux and so the Wess-Zumino part of the brane action does not give rise to a Chern-Simons coupling for our $U(N_f)$ gauge field. This argument was first made in~\cite{Jensen:2010em} (see also \cite{Hung:2009qk}) and works for Abelian as well as non-Abelian gauge fields.
Then up to some constant the action for the probe $D3'$-branes will be the non-Abelian DBI-action~\cite{Myers:1999ps}, which
after exploiting our symmetries\footnote{See section 3 of~\cite{Ammon:2009fe} for details of the analogous computation for probe $D7$ branes.} amounts to
\begin{equation}\label{eq:D3'action}
 S_{D3'}=-T_{D3}N_f Str\int dt dx dz d\xi \sqrt{-det(g_{ab}+(2\pi \alpha')F_{ab})} \, ,
\end{equation}
with $F=dA+A\wedge A$, and the symmetrized trace $Str$ is taken over $U(N_f)$ representation matrices (we have renamed the radial coordinate $u\to z$ for convenience). Here we choose the metric and gauge field to be indepenent of the $S^1$-coordinate $\xi$.
After expanding the square root in \eqref{eq:D3'action} in small field strengths $F$ the leading contribution 
%$S_{\text{}} \propto \int dt dx dz \sqrt{-\det(g_{ab})}$ 
is extremized by the $AdS_3$ black hole background. The subleading contribution merely has the form of the Yang-Mills action which we already advertised in equation \eqref{eq:action}. It should be noted here that the non-Abelian DBI action is only valid up to fourth order in field strengths~\cite{Tseytlin:1997csa,Hashimoto:1997gm}, and it has been shown to disagree with corresponding string scattering amplitude computations beyond this order. However, in the present paper we are only interested in the leading (quadratic in field strengths) order.

The field theory which is holographically dual to this gravity setup is known and has originally been studied in~\cite{Constable:2002xt}, see also~\cite{Jensen:2010em,Hung:2009qk}. This dual field theory is $U(N_f)\times U(N)\, \mathcal{N}=4$ SYM theory coupled to a bifundamental hypermultiplet along the $(1+1)$-dimensional defect. In particular there is a dynamical gauge field $\mathfrak{A}$ living on the $(1+1)$-dimensional defect. This dynamical gauge field is dual to the dynamical bulk $U(N_f)$ gauge field $A$ on the stack of $N_f$ probe branes. Therefore the setup presented in this subsection provides an example of the dynamical boundary gauge fields which we want to study in this work.
It should be noted that our dynamical boundary gauge field $\mathfrak{A}$ (dual to the bulk gauge field $A$) is still a gauge-independent operator under the original "color" group $U(N\to\infty)$. It is only dynamical and hence gauge-dependent under the additional "flavor" $U(N_f)$ which we have introduced.

%-------------------------------------------------
\subsection{Yang-Mills-Action in AdS$_3$}
In the previous subsection we have illustrated how the Yang-Mills action \eqref{eq:action} may arise from a certain string construction. In this subsection we focus on this Yang-Mills action for a non-Abelian gauge field $A$ in $AdS_3$. The neutral AdS$_3$ black hole background in Poincar\'{e} coordinates is
given by
\begin{equation}
ds^2=\frac{L^2}{z^2}(-f(z)dt^2+dx^2+\frac{dz^2}{f(z)}) \, ,
\end{equation}
in which $f(z)=1-z^2$, $z=r_+/r$, and $r_+$ is the horizon of the black hole.
The black hole temperature is given by
\begin{equation}
T=\frac{r_+}{2 \pi } \, .
\end{equation}

We are free to set $L=1$ and $r_+=1$.
To this background, we add $SU(2)$ gauge fields $\mathcal{A}_\mu^a$, with
$a=1,2,3$.
We split this bulk gauge field $\mathcal{A}_\mu^a$ in background fields
$A_\mu^a$, and fluctuations  $a_\mu^a$
around this background. Our Ansatz for the background fields is
\begin{equation}\label{eq:pWaveAnsatz}
A = \Phi(z) dt \tau^3 + W(z) dx \tau^1 \, ,
\end{equation}
where $\tau^a$ are the generators of the $SU(2)$. For our fluctuations we choose
the gauge $a^b_z\equiv0$ but leave them otherwise unrestricted for now.
The field strength is
${\cal F}_{\mu \nu}^a = \partial_\mu {\cal A}^a_\nu -
\partial_\nu {\cal A}^a_\mu +g_{YM} \epsilon^{abc} {\cal A}^b_\mu {\cal A}^c_\nu$.
For simplicity we set $g_{YM}\equiv 1$ in the remainder of this work. Note
that the value of $g_{YM}$ would be important if we were to relax the probe limit,
comparing the gravitational coupling to the gauge coupling.

The equations of motion for the p-wave background fields $\Phi$ and $W$ are
\begin{eqnarray}\label{eq:backgroundEOM}
0 &=& \partial_z(zf\partial_zW)+\frac{z}{f}\Phi^2W\, , \label{eq:wEq} \\
0 &=& \partial_z(z\partial_z\Phi)-\frac{z}{f}\Phi W^2\, .\label{eq:phiEq}
\end{eqnarray}

%----------------------------------
\subsection{Alternative Boundary Conditions}
\label{sec:altBC}
The boundary condition at the horizon is $\Phi(z=1)=0$,
which is to make $\Phi dt$ well-defined at $z=1$, while
$W$ should be regular at the horizon. The horizon expansion is
\begin{eqnarray}\label{eq:horizonExpBackground}
\Phi & = & -\Phi_1 (z-1)
+\frac{1}{4}\left(2 \Phi_1+{W_0}^2 \Phi_1\right)(z-1)^2
+ {\cal O}((z-1)^3) \, ,\\
W & = & W_0 + {\cal O}((z-1)^3)
\, ,
\end{eqnarray}
This expansion has the same structure
as in the AdS$_5$/CFT$_4$ and AdS$_4$/CFT$_3$ cases.
In contrast to that, at the boundary, the story is different. The
asymptotic behavior of $W$ and $\Phi$ at the boundary $z=0$ is
\begin{eqnarray}\label{eq:wPhi}
W &=& -\mathfrak{J}_x\ln z+\langle\mathfrak{A}_x\rangle\, , \nonumber \\
\Phi &=& -\mathfrak{J}_t \ln z+ \langle\mathfrak{A}_t\rangle\, ,
\end{eqnarray}
where $\mathfrak{J}_t=:\rho$ and $\mathfrak{J}_x$ are the sources,
and $\langle\mathfrak{A}_t\rangle=:\mu$, as well as $\langle\mathfrak{A}_x\rangle$
are the vacuum expectation values\footnote{Note that we have supressed gauge indices $a, b, \dots$
since according to our Ansatz \eqref{eq:pWaveAnsatz} in the background the
$a=3$ index always appears with the spacetime index $\mu=t$
and $a=1$ appears always with $\mu=x$.}
of the operator dual to the bulk background field $A_\mu^a$.
Note that we would get a different boundary behavior for the gauge fields had we
allowed a Chern-Simons term in our action~\eqref{eq:action}. In that case
the asymptotics depend on the Chern-Simons level, see for example~\cite{Jensen:2010em}.

In \eqref{eq:wPhi} we now encounter the alternative quantization
mentioned above, which corresponds to the choice of Neumann boundary
conditions~\cite{Domenech:2010nf,Montull:2012fy,Silva:2011zz,Montull:2011im}.
In AdS$_4$ for example, we would be free to
interpret the leading term of $\Phi$ as the source (chemical potential),
and the subleading term as the expectation value of the dual current operator (charge density),
or the other way around. However, this
is not possible in AdS$_3$ ~\cite{Jensen:2010em, Marolf:2006nd}. 
We have to choose the Neumann boundary condition. One may be concerned that~\cite{Jensen:2010em, Marolf:2006nd} consider Abelian gauge fields for the most part. But it has been noted in~\cite{Marolf:2006nd} and earlier in~\cite{Witten:2003ya} that the asymptotics and hence also the boundary conditions are determined by the linear theory. Our non-Abelian terms contribute to non-linear (interaction) terms.\footnote{As a concrete example, boundary conditions have been discussed in~\cite{Marolf:2006nd} for $SO(8)$ non-Abelian gauge fields of $AdS_4$ supergravity.} Therefore our non-Abelian gauge fields in $AdS_3$ should only satisfy the Neumann boundary condition, just like their Abelian counterparts.

So compared to the higher dimensional cases, we observe two changes:
First, the constant term is not the leading contribution anymore. A
logarithmic term appears and contributes to leading order.
Second, as explained in~\cite{Jensen:2010em, Marolf:2006nd}, the leading (now logarithmic)
term has to be interpreted as a source and corresponds to a (non-dynamical)
boundary current. Therefore, the source in this
system is the boundary current $\mathfrak{J}$. The subleading (constant) term
is the expectation value of the operator dual to the field $A$. We denote
this operator by $\mathfrak{A}$. According to the analysis in~\cite{Jensen:2010em, Marolf:2006nd},
this operator $\mathfrak{A}$ has to be interpreted as a gauge field
living on the AdS$_3$-boundary. So in summary, our bulk gauge field
$A$ is dual to a boundary gauge field $\mathfrak{A}$. This boundary
gauge field $\mathfrak{A}$ is sourced by the current $\mathfrak{J}$.

Thus our fixed charge density $\rho = \mathfrak{J}_t$ sources the dynamical
gauge field $\mu=\langle\mathfrak{A}_t\rangle$. 
This particular component $\langle\mathfrak{A}_t\rangle$ can be interpreted as a chemical potential (which does not act as a source, but is determined by the equations of motion).
Meanwhile $\langle\mathfrak{A}_x\rangle$ is the order parameter of the symmetry-breaking,
while $\mathfrak{J}_x$ is the source term which we demand to
vanish, for the breaking to be spontaneous. Our operator $\mathfrak{A}_x$ is the spatial component of a dynamical boundary gauge field, which acquires an expectation value. In the context of thermodynamics
our alternative setup has a simple interpretation: since we are working
at fixed charge density, this is the canonical ensemble (provided
we add an appropriate boundary term to the action~\eqref{eq:action} 
as discussed below).

%----------------------------------
\subsection{Symmetry-Breaking Solution and Free Energy}
For simplicity we choose to work in the probe limit, i.e. the gauge field does
not backreact on the AdS$_3$ metric.
After solving the equations of motion \eqref{eq:backgroundEOM} with boundary conditions \eqref{eq:horizonExpBackground}, the order parameter
$\langle\mathfrak{A}_x\rangle$ is plotted versus the temperature in figure \ref{fig:condensate}, both in units of $T_c$. 
Note that in the canonical ensemble $T_c$ is 
determined by the only other dimensionful quantity in the problem, namely $\rho$, see~\cite{Hartnoll:2008vx,Hartnoll:2008kx}. So we actually plot 
$\langle\mathfrak{A}_x\rangle$ rescaled by $(\rho 2\pi \rho_c)$ versus the temperature as measured by $T/T_c=\rho_c/\rho$.
\FIGURE{
\includegraphics{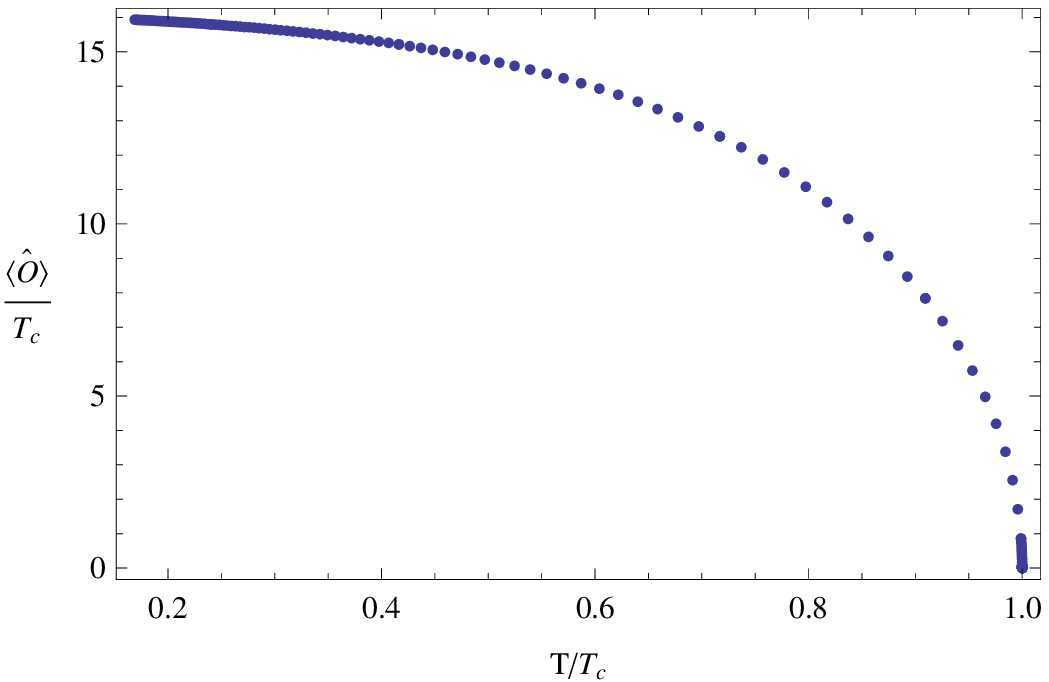}
\caption{
Condensate $\langle\hat{\mathcal{O}}\rangle/T_c$ =  $\langle\mathfrak{A}_x\rangle/(\rho 2\pi \rho_c)$ versus temperature
$T/T_c = \rho_c/\rho$; the critical temperature
is given by $T_c=0.046\rho$. \label{fig:condensate} 
}
}

Near the critical temperature $T_c$, the 
condensate $\langle\mathfrak{A}_x\rangle$ behaves like
\begin{eqnarray}
 \langle\mathfrak{A}_x\rangle \approx 25.5515 T_c
\sqrt{1-T/T_c}.
\end{eqnarray}
The critical exponent is $1/2$ which is identical to the one in mean
field theory. Note that this solution also breaks parity in $(1+1)$ dimensions.
In any odd number of spatial dimensions parity is defined as inversion
of only one of the spatial coordinates. In our setup this would be inversion
of the one spatial coordinate $x$. Our order parameter $\langle\mathfrak{A}_x\rangle$
is a vector pointing along this $x$-direction (or against it depending on the
sign conventions). Therefore parity symmetry, the invariance of our system under
inversion of the $x$-coordinate, is broken spontaneously.

In the following we will compute the free energy
difference between the normal state and the broken
state in order to verify that the broken phase is preferred at low temperatures.
With the Ansatz \eqref{eq:pWaveAnsatz}, we can rewrite
\begin{equation}
\frac{1}{2}\sqrt{-g}\mathcal{L}= -\frac{1}{2}\sqrt{-g} F^a_{\mu \nu}F^{a \mu \nu}=-z f (\partial_z W)^2+\frac{z}{f} \Phi^2 W^2+z(\partial_z W)^2 \, .
\end{equation}

Then the on-shell action
\begin{equation}
\begin{split}\label{eq:Sos}
S_{\text{os}}=\int \sqrt{-g}\mathcal{L} d^3x  =
2 \lim\limits_{\epsilon\to 0}\int dt \,dx\, \left(-z f W \partial_z W + z \Phi \partial_z\Phi\right)_{z=\epsilon} - 2 \int dz\, dt\, dx\, \frac{z}{f} \Phi^2 W^2 \, ,\\
=2 \int dt \,dx \,\left(-\mu \rho + \rho^2 \log(z)\right) - 2 \int dz\, dt\, dx\, \frac{z}{f} \Phi^2 W^2
\, ,
\end{split}
\end{equation}
can be obtained by integrating our action~\eqref{eq:action} by parts, and using the equations of motion~\eqref{eq:backgroundEOM} in the form
\begin{equation}
-\int dz[zf(\partial_z W)^2]=-z f W  \partial_z W |_{z=\epsilon}-\int dz \frac{z}{f}\Phi^2 W^2\, ,
\end{equation}
and
\begin{equation}
\int dz [z(\partial_z \Phi)^2 ]=  z \Phi \partial_z \Phi|_{z=\epsilon}-\int dz \frac{z}{f}\Phi^2 W^2\, .
\end{equation}
In order to place the dual field theory in the canonical ensemble we need to
add a boundary term~\cite{Marolf:2006nd, Jensen:2010em}
\begin{equation}\label{eq:Sbdy}
S_{\text{bdy}} =  -2 \int dt\, dx\, \left(\sqrt{-g} A_\mu F^{z\mu}\right)_{z=0}
= 2 \int dt dx (- z f W'W + z \Phi' \Phi )_{z=0}
 \, ,
\end{equation}
where the prime denotes a derivative with respect to the radial coordinate $z$.
Note that this contribution~\eqref{eq:Sbdy} is identical to a part of the 
on-shell action~\eqref{eq:Sos}.
Now we have to add a counterterm
\begin{equation}
S_{\text{counter}} = 4 \int dt \,dx \, \sqrt{-h} F_{z\mu} F^{z\mu} \log(z)
= - 4 \int dt \, dx \, \rho^2 \log(z) \, ,
\end{equation}
in order to cancel the logarithmic divergences in~\eqref{eq:Sos} and~\eqref{eq:Sbdy}.
Here $h$ is the metric induced on the boundary.
Hence we obtain the renormalized on-shell action
\begin{equation}\label{eq:Sren}
S_{\text{ren}} = 
S_{\text{os}}+S_{\text{bdy}} + S_{\text{counter}}
= -4 \int dt \int dx \rho \mu- 2\int d x^3 \frac{z}{f}\Phi^2 W^2\,,
\end{equation}
where we have used that the source has to vanish, i.e. ${\mathfrak{J}_x}=0$ 
is required, for spontaneous symmetry-breaking.
Finally the free energy density $\mathcal{F}$ is given by the Euclidean continuation
of $S_{\text{ren}}$ as
\begin{equation}
\mathcal{F}= -T S_{\text{ren}} =4 \mu \rho + 2 \int dz \frac{z\Phi^2 W^2}{f} \, .
\end{equation}
Note that the free energy of the normal state in our conventions is always zero,
because in that phase the general gauge field solution is $A_t^3= -\rho \log(z)+C_0$ and the constant $C_0$ has to vanish for $A_t^3(z)$ to be well-defined at the horizon.
This situation would change if we were to consider backreaction; then 
we would have finite $\rho$ and $\mu$.

The free energy difference between the broken state
and the normal state is plotted in figure \ref{fig:freeenergy}, from which we can see
that the state with a condensation is thermodynamically preferred
when $T<T_c$. The condensate vanishes with a square root behavior.

Thus our first result is that there is a second order
phase transition for the Yang-Mills theory~\eqref{eq:action} in the AdS$_3$
black hole background at the critical temperature $T_c$. Our
broken phase develops a (parity-breaking) vector condensate, which corresponds to
vector hair of the AdS$_3$ black hole.
\FIGURE{
\includegraphics{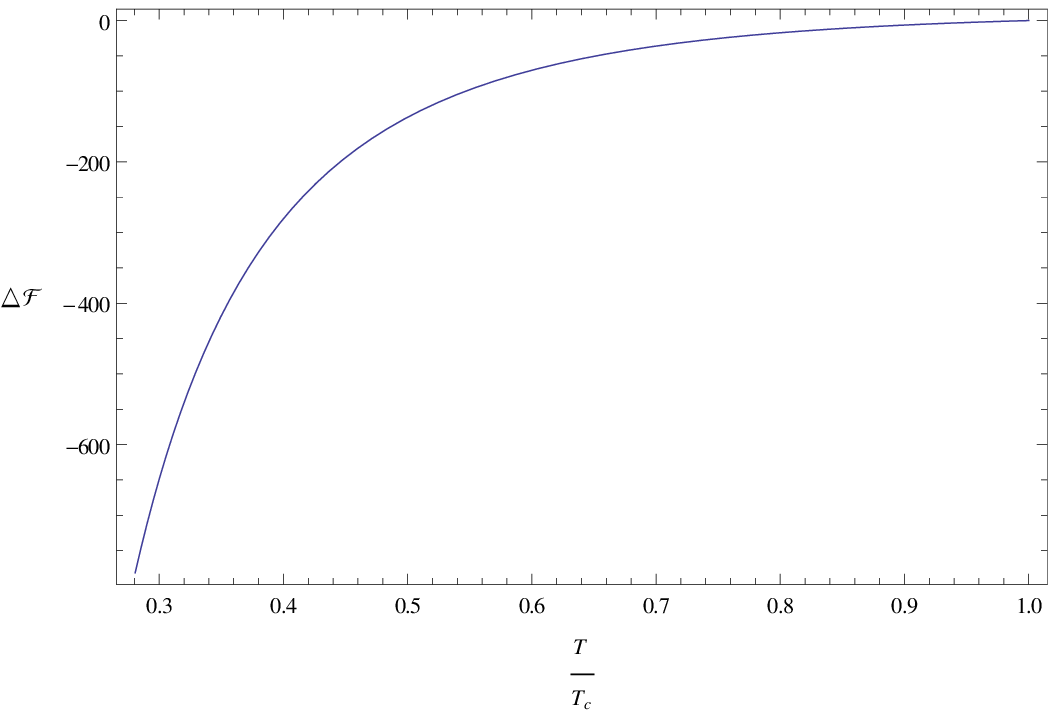}
\caption{The
free energy difference between the normal and broken phase. \label{fig:freeenergy} }
}

The p-wave solution we have presented here is different from the
one discussed in \cite{Gubser:2008wv}. Our lowest energy solution for
the field $W(z)$ has one node at some $z=z*$,
while the p-wave in AdS$_4$ has no nodes. 
One may argue that more nodes
should correspond to a higher energy for the solution~\cite{Gubser:2008wv}. According to that reasoning 
the zero node solution should always be lowest in energy. However, 
in our AdS$_3$ case
the zero node solution
coincides with the solution which has $W(z)\equiv 0$. Or in other words:
there is no symmetry-breaking solution with zero nodes in our setup.

%%%%%%%%%%%%%%%%%%%%%%%%%%%%%%%% RESULTS
\section{Fluctuations} \label{sec:fluctuations}
In this work we will not be able to address all the field theory features presented in
section \ref{sec:hydro} at once. Instead we are going to pick
selected modes in order to investigate the
dynamics of the full system in the normal phase and in the phase with
spontaneously broken symmetry, i.e. the broken phase. Our goal is
to show that there are non-trivial dynamical low-energy modes near
the phase transition in the broken phase of this system.

\subsection{Fluctuation Equations of Motion}
We start by writing down the full set of equations of motion
after the gauge choice $a_z\equiv 0$ is made
\begin{eqnarray} \label{eq:eoms}
0 & = & {a_t^1}'' +\frac{1}{z} {a_t^1}'-\frac{k^2}{f} a_t^1
-\frac{k\omega a_x^1-(i k a_x^2+W a_x^3)\Phi}{f}\, ,\nonumber\\
0 & = & {a_x^1}'' +\left( \frac{1}{z}+\frac{f'}{f}\right) {a_x^1}'
+\frac{\omega^2+\Phi^2}{f^2} {a_x^1}
+\frac{\omega k {a_t^1}+\left(
-i(k a_t^2+2\omega a_x^2) + 2 a_t^3 W
\right) \Phi}{f^2}\, ,\nonumber\\
0 & = & {a_t^2}'' + \frac{1}{z} {a_t^2}'- \frac{k^2 + W^2}{f} {a_t^2}
- \frac{k\omega a_x^2 + i (2k a_t^3 W +\omega a_x^3 W+k a_x^1 \Phi)}{f} \, ,\nonumber\\
0 & = & {a_x^2}'' + (\frac{1}{z}+\frac{f'}{f} ){a_x^2}'+\frac{\omega^2+\Phi^2}{f^2} a_x^2
+\frac{k\omega a_t^2 + i(\omega a_t^3 W+(k a_t^1+2\omega a_x^1)\Phi)}{f^2} \, ,\nonumber\\
0 & = & {a_t^3}'' + \frac{1}{z}{a_t^3}'-\frac{k^2+W^2}{f}a_t^3
-\frac{k\omega a_x^3 + W (-i(2k a_t^2 +\omega a_x^2)+2 a_x^1 \Phi)}{f}  \, ,\nonumber\\
0 & = & {a_x^3}'' +(\frac{1}{z}+\frac{f'}{f}) {a_x^3}' + \frac{\omega^2}{f^2} a_x^3
+ \frac{k \omega a_t^3 -i \omega a_t^2 W - a_t^1 W \Phi}{f^2}   \, .
\end{eqnarray}

Variation of the action with respect to the fields $a_z$ yields the
following constraints
\begin{eqnarray}\label{eq:constraints}
0 & = & -i \omega  {a_t^1}'-\Phi {a_t^2}'-i k f  {a_x^1}'+{a_t^2} \Phi' \, , \nonumber\\
0 & = & \Phi {a_t^1}'-i \omega  {a_t^2}'-f \left(i k {a_x^2}'-W {a_x^3}'+{a_x^3} W'\right)-{a_t^1} \Phi ' \, , \nonumber \\
0 & = & -i \omega  {a_t^3}'-f \left(W {a_x^2}'+i k {a_x^3}'-{a_x^2} W'\right) \, .
\end{eqnarray}

Expanding the fluctuations~\eqref{eq:eoms} around the AdS-boundary, we get
\begin{eqnarray}\label{eq:boundaryFlucs}
a_t^a & = & {a_t^a}^{(\text{source})} \log(1/z) + {a_t^a}^{(\text{vev})}+\dots \, , \\
a_x^a & = & {a_x^a}^{(\text{source})} \log(1/z) + {a_x^a}^{(\text{vev})}+\dots \, ,
\end{eqnarray}
where $a=1,2,3$ is the gauge index, (source) denotes the source, and
(vev) denotes the vacuum expectation value. We follow the same
logic which was applied for our background fields $A$ below \eqref{eq:wPhi}.
So the coefficient of the logarithmic term in \eqref{eq:boundaryFlucs}
is interpreted as the boundary theory current $\mathfrak{j}_\mu^b = {a_\mu^b}^{(\text{source})} $ which acts as a source.
The constant term is interpreted as the vacuum expectation value of
a boundary theory gauge field $\langle\mathfrak{a}_\mu^b\rangle = {a_\mu^b}^{(\text{vev})} $.

Near the horizon these fluctuations behave as
\begin{eqnarray}\label{eq:horizonFlucs}
a_t^a & = & (1-z)^{-i\frac{\omega}{2}}
\left (\hphantom{{a_t^a}^{(\text{H})}+} \mathcal{O}(1-z)\right)  \, , \\
a_x^a & = & (1-z)^{-i\frac{\omega}{2}}\left ({a_x^a}^{(\text{H})} +\mathcal{O}(1-z)\right)  \, .
\end{eqnarray}
All higher coefficients depend recursively on the three initial horizon 
values ${a_x^1}^{(\text{H})},\,
{a_x^2}^{(\text{H})}$ and ${a_x^3}^{(\text{H})}$ of the spatial components.

\subsection{Analytical Results}\label{sec:methods}
\paragraph{Gauge-invariant field combinations}{
Above we have constructed a ground state which
breaks the $SU(2)$ gauge symmetry explicitly to $U(1)_3$, which is
broken spontaneously by the same ground state.
Therefore our ground state does not have any
gauge symmetry left. However, the Lagrangian of our theory still enjoys
the full $SU(2)$ gauge symmetry. Thus, while the $SU(2)$-symmetry is completely broken for our background fields which describe the ground state, the fluctuations still transform under a full $SU(2)$ gauge symmetry in the following way
\begin{equation}
\delta a_\mu^a = \partial \lambda^a + \epsilon^{abc} A_\mu^b \lambda^c\, ,
\end{equation}
with the transformation parameters $\lambda^a(t,x)$ where $a=1,\,2,\,3$.
These $\lambda^a$ do not depend on the radial coordinate $z$ because
we have chosen to work in the axial gauge $\mathcal{A}_z\equiv 0$ earlier.

First, we note that there are three pure gauge solutions for the fields
$(a_t^1, a_x^1, a_t^2, a_x^2, a_t^3, a_x^3)$, which in this notation read
\begin{equation}
\label{eq:pureGauge}
\begin{array}{c c c c c c c c c}
(I)&:&(&-i\omega,&ik,& \Phi,&0,&0,&0) \, ,\\
(II)&:&(&-\Phi,&0,&-i\omega,&ik,&0,&W) \, ,\\
(III)&:&(&0,&0,&0,&-W,&-i\omega,&ik) \, .
\end{array}
\end{equation}
We have labeled these solutions $(I), (II), (III)$ for later convenience.

We find the following gauge-invariant bulk field combination:
\begin{equation}\label{eq:aHat}
\hat a_x^3 = a_x^3 + \frac{k}{\omega} a_t^3 + W \frac{\Phi a_t^1+i\omega a_t^2}{\Phi^2-\omega^2} \, .
\end{equation}
At vanishing momentum this expression reduces
to the gauge-invariant combination considered by Gubser \& Pufu in \cite{Gubser:2008wv} .

Of course we can build more gauge-invariant combinations such as
\begin{equation}
\hat a_x^2 = a_x^2 + \frac{i}{\omega W}(W^2-k^2) a_t^3 - \frac{i k}{W} a_x^3 \, .\nonumber
\end{equation}
}
These are discussed in appendix~\ref{sec:covariantFields}.

In the most general case below $T_c$, all six fluctuations are coupled through the condensates, momenta and frequency terms in the equations of motion~\eqref{eq:eoms}. Therefore all the two-point functions will have the same poles. For a more detailed discussion see
\cite{Amado:2009ts,Kaminski:2009dh}. We will be interested in finding these common
poles in order to identify low-energy modes. These are the poles at low frequency
and small momentum which govern the long-time behavior of our system.
By one of our inital assumptions we identify the poles in our two-point functions
with the quasinormal modes of the gauge-invariant bulk field combinations
$\hat a^b_\mu$ defined above. Our quasinormal modes can in turn be found
by a method which we outline below. In the same way
in which all correlators contain the same poles, the bulk fields will all have
identical quasinormal modes in the fully coupled system of six equations of
motion \eqref{eq:eoms}.

\paragraph{Gauge-covariant field combinations}
In the normal phase the symmetry is only broken partly by the ground state.
There is a $U(1)_3$ remaining intact. Therefore the gauge transformation
for the fluctuations in this case amounts to
\begin{equation}
\delta a_\mu^a = \partial \lambda^a + \epsilon^{abc} (A_\mu^b \lambda^c
+a_\mu^b \Lambda^3\delta^c_3)\, ,
\end{equation}
where $\Lambda^3$ is the gauge transformation parameter of the
remaining $U(1)_3$ in the background.

It is not possible to form gauge-invariant combinations of the fluctuations
$a_\mu^b$ as we did before in the broken phase. However, we can
form gauge-covariant combinations, i.e. combinations which will have
a well-defined homogeneous transformation under the gauge-symmetries.

We consider the following gauge-covariant field combinations
\begin{eqnarray}
e_L^{+} &=& (\Phi(z) - \omega) e_{x}^{+} -  k e_{t}^{+}\, , \\
e_L^{-} &=& (\Phi(z) + \omega) e_{x}^{-} +  k e_{t}^{-}\, , \\
e_3 &=& k a_t^3 + \omega a_x^3\, ,
\end{eqnarray}
where $e_t^{\pm} =a_t^1 \pm  i a_t^2 $, $e_x^{\pm} =a_x^1 \pm i a_x^2$.
Note that $e_3$ is gauge-invariant and thus corresponds to a
physically meaningful operator on the boundary by our initial assumptions.
Details are discussed in appendix~\ref{sec:covariantFields}, 
see also~\cite{Ammon:2011je}.

\paragraph{Determinant Method \& Quasinormal Modes}
Quasinormal modes are those bulk field solutions at a particular
$\omega=\omega_{n,\text{QNM}}$ which are infalling at the horizon
and satisfy a Dirichlet condition at the AdS-boundary; we choose them
to vanish at the AdS-boundary. In a coupled system it is not a trivial task
to identify the relevant field combinations which have to vanish for a
quasinormal mode to be present. A systematic method has been
developed in~\cite{Amado:2009ts,Kaminski:2009dh}. However,
we will see that this method needs to be extended in our case
at hand, because there are logarithmic sources.

Let us solve the equations of motion numerically, for example with the horizon values
${a_x^1}^{(\text{H})} = -1$, ${a_x^2}^{(\text{H})} = 1$, and ${a_x^3}^{(\text{H})} = 1$,
see~\eqref{eq:horizonFlucs}. We label this solution $(IV)$. Then
we solve the equations of motion again using different horizon values
${a_x^1}^{(\text{H})} = 1$, ${a_x^2}^{(\text{H})} = -1$, and ${a_x^3}^{(\text{H})} = 1$. This solution is linearly independent
from the first one and we label it $(V)$. There is a third linearly independent
solution which we find using the values ${a_x^1}^{(\text{H})} = 1$, ${a_x^2}^{(\text{H})}= 1$, and ${a_x^3}^{(\text{H})} = -1$,
and we label it $(VI)$. It was shown in \cite{Amado:2009ts,Kaminski:2009dh}, that these
three solutions can be combined in a matrix together with the three pure
gauge solutions given in \eqref{eq:pureGauge}. The determinant of this
matrix, evaluated at the boundary $z=0$ was shown to yield locations
$\omega^{(n)}_{QNM}(k)$ of quasinormal modes of the coupled system.
Quasinormal modes can be found by considering the following
determinant
\begin{equation}
{\cal D} = \det \left (
\begin{array}{c c c c c c}
-i\omega & ik & \Phi & 0 & 0 & 0\\
-\Phi & 0 & -i\omega & ik &0 & W\\
0 & 0 & 0 &-W & -i\omega & ik \\
{{a_t^1}^{(IV)}} & {{a_x^1}^{(IV)}} & {{a_t^2}^{(IV)}} &{{a_x^2}^{(IV)}} & {{a_t^3}^{(IV)}} & {{a_x^3}^{(IV)}} \\
{{a_t^1}^{(V)}} & {{a_x^1}^{(V)}} & {{a_t^2}^{(V)}} & {{a_x^2}^{(V)}} & {{a_t^3}^{(V)}} & {{a_x^3}^{(V)}} \\
{{a_t^1}^{(VI)}} & {{a_x^1}^{(VI)}} & {{a_t^2}^{(VI)}} & {{a_x^2}^{(VI)}} & {{a_t^3}^{(VI)}} & {{a_x^3}^{(VI)}} \\
\end{array}
\right ) \, .
\end{equation}
In~\cite{Amado:2009ts,Kaminski:2009dh} this kind of determinant was required to vanish in the following way:
\begin{equation}\label{eq:oldDeterminantMethod}
0=\lim\limits_{z\to 0} \,{\cal D}\, .
\end{equation}
This procedure  essentially sets the relevant sources to zero if we work with bulk fields which have
a constant as the source term, and no more divergent terms. 
However, in our AdS$_3$ case we have
logarithms. Therefore the recipe has to be changed in order to extract the
sources which now hide in the logarithmic terms. One sensible prescription
appears to be
\begin{equation} \label{eq:newDeterminantMethod}
0=\lim\limits_{z\to 0} z\,  \partial_z \,{\cal D}\, .
\end{equation}
Whenever we set a fluctuation consistently to zero, we should eliminate
the corresponding column and one row of solutions from this determinant.
For example this prescription yields exactly the gauge-invariant combination
$\hat{a}_x^3$ defined in~\eqref{eq:aHat}, when we set $a_x^1=0=a_x^2$ at
$\Phi,\,W,\neq0$, but $k=0$.

Let us work through this example: we have two pure gauge
solutions $a = (-i\omega, \Phi, 0)$ and $a = (-\Phi, - i\omega, W)$,
and one non-trivial solution, which depends on one parameter, namely
the source of $a_x^3$ at the horizon.
So this gives a solution matrix
\begin{equation}
H =\left(
\begin{array}{c c c}
-i\omega & \Phi & 0 \\
-\Phi & - i\omega & W \\
a_t^1 & a_t^2 & a_x^3
\end{array}
\right) \, .
\end{equation}
Now we compute the determinant
$\mathcal{D} = \det H =  a_x^3 - W (\Phi a_t^1 + i\omega a_t^2)/(\omega^2-\Phi^2)$,
and require it to vanish in general. This yields exactly the gauge-invariant combination
$\hat{a}_x^3$ which was found in the analogous AdS$_4$-system~\cite{Gubser:2008wv}.
But there is a crucial difference now. In AdS$_4$ we simply had to take
the prescription \eqref{eq:oldDeterminantMethod} because the sources were
the asymptotic values of the fields, and $\hat{a}_x^3$ was just a linear combination of these. Luckily the latter is still true in AdS$_3$, but now we have to extract
the logarithm coefficients. This leads to the adjusted prescription \eqref{eq:newDeterminantMethod}.
The reason why this works here is that for $k=0$ we can reduce~\eqref{eq:eoms} 
to three second order
equations of motion for $a_t^1$, $a_t^2$, and $a_x^3$. These three equations are coupled. But gauge symmetry restricts them, so that they can be reduced to
one single equation. In other words:
When we work out all of the 9 equations of motion (not choosing a gauge yet), then we see that there are 2 constraint equations from \eqref{eq:constraints} which relate
$a_t^1$ to $a_t^2$ and $a_x^3$. Then the counting is: 3 equations of motion with
each 2 boundary conditions give 6 boundary conditions. Choosing the infalling solution for each of the 3 fields there are 3 boundary conditions left to be fixed. Now we can use our 2 contraints, which leaves merely one solitary boundary condition. This means that there is only one relevant field $\hat{a}_x^3$ and our one boundary condition determines the normalization of this field.

As a second example, we consider the normal phase with $W=0$. 
Here the adjusted method~\eqref{eq:newDeterminantMethod} yields accordingly
$e_3 = \omega a_x^3 + k a_t^3$ at $a_{t,x}^1=0= a_{t,x}^2$ and $\Phi\neq 0$,
and correctly requires the leading logarithmic term in $e_3$ to vanish.
However, for the genuinely coupled cases the prescription~\eqref{eq:newDeterminantMethod} may have to be modified.\footnote{
Equivalently, at least in the normal phase, one can carry out an analysis
formally similar to~\cite{Hung:2009qk}.} In particular the gauge-covariant
gauge fields will not necessarily be linear combinations of the individual
fields $a^a_\mu$ anymore. We postpone their consideration to future work.

Numerically we can not take the limit $z\to 0$ exactly. Therefore we have to
choose the same cut-off $\Lambda = 10^{-9}$ which we had chosen when
computing our background data $\Phi$ and $W$. See~\cite{Amado:2009ts,Kaminski:2009dh} for a more detailed discussion.

\subsection{Numerical Results} \label{sec:numerics}

Here we will simplify our lifes and consider only particular gauge-invariant 
field combinations which
decouple from all others. Such field combinations are completely determined by
requiring the infalling boundary condition at the horizon, and by giving a single second
boundary condition fixing the normalization of the solution. For a discussion
of how to find these decoupled solutions see section~\ref{sec:methods}
and the appendix~\ref{sec:covariantFields}. Note that the frequency $\omega$
in the following plots should be understood as the dimensionless quantity
$\omega\to\omega/(2\pi T)$, and similarly $k\to k/(2\pi T)$.

\paragraph{Normal phase at vanishing spatial momentum $k=0$}
We solve the equations of motion~\eqref{eq:eoms} in the normal phase
at finite chemical potential induced by $\Phi\neq 0$ at $\omega\neq 0$,
but at $W=0$ and $k=0$. In this case the last equation in~\eqref{eq:eoms}
can be consistently decoupled from the others.
Therefore we consider that single equation for the fluctuation $a_x^3$, setting all other fluctuations to zero
\begin{equation}\label{eq:ax3Only}
0 = {a_x^3}'' + \frac{1-3z^2}{z(1-z^2)}{a_x^3}' +\frac{\omega^2}{(1-z^2)^2} {a_x^3} \, .
\end{equation}
Note that this equation does not depend on the background field $\Phi$
in other words the dual operator is not charged under $U(1)_3$. 
We consider the Green's function of the dynamical gauge field
combination ${\mathfrak{a}}_x^3$ in terms of the gauge-invariant
electric field ${\mathfrak{e}}_x^3 = \omega {\mathfrak{a}}_x^3$.
The straightforward recipe for correlation functions (up to contact terms) can
be applied here as shown in~\cite{Jensen:2010em}, to yield
\begin{equation}
G^R_{\mathfrak{e}_x^3\mathfrak{e}_x^3} 
=\omega^2 G^R_{{\mathfrak{a}}_x^3{\mathfrak{a}}_x^3} 
= \omega^2 \frac{(\text{vev})}{(\text{source}) }
=  \omega^2\frac{{a_x^3}-\log (z) z \partial_z {a_x^3}}{z \partial_z {a_x^3}} \, .
\end{equation}
Note that we are consequently ignoring numerical factors for the sake of a 
simple presentation here.
Solving the equation of motion~\eqref{eq:ax3Only}, we see that the imaginary part of the Green's function here asymptotes to a constant at large frequencies and has a pole at $\omega=0$
independent of charge density.

\paragraph{Normal phase at non-vanishing spatial momentum $k\not=0$}
We solve the equations of motion~\eqref{eq:eoms} in the normal phase with $W=0$
at finite momentum $k\neq0$ and at finite chemical potential
induced by $\Phi\neq 0$, while  $\omega\neq 0$.
Now the last two equations in~\eqref{eq:eoms}
decouple from the others. Therefore we consider only the two coupled equations
for the two fluctuations $a_t^3$ and $a_x^3$. These can be combined
into $e_3$ as discussed above. For simplicity we set all other fluctuations to zero and
only study $e_3$ here. There is one pure gauge solution $a=(-i\omega,i k)$, and
one non-trivial solution determined by a single parameter, which we choose to be
${a_x^3}(z=z_H)$ at the horizon $z_H=1$.
Similar to the previous case we define the
Green's function 
\begin{equation}
G^R_{\mathfrak{e}_3\mathfrak{e}_3} 
= \frac{(\text{vev})}{(\text{source}) }
= \frac{{e_3}-\log (z) z \partial_z {e_3}}{z \partial_z {e_3}} \, .
\end{equation}
Note that $e_L^\pm$ satisfy coupled equations and
would require a more careful treatment. For $e_3$ however we numerically find
a linear dispersion relation
\begin{equation}
\omega = \pm k \, ,
\end{equation}
as seen in figure~\ref{fig:normalE3E3}; in formal agreement with~\cite{Hung:2009qk}.
However, note that the authors of~\cite{Hung:2009qk} choose to interpret
their correlation functions as current-current correlation functions. In contrast
to that we interpret our two-point functions according to the alternative setup
as gauge field correlators. In this context it is reassuring that the
gapless mode propagates with the speed of light $c=1$ at finite
temperature and charge density. It behaves just like a free photon.
\FIGURE{
\includegraphics{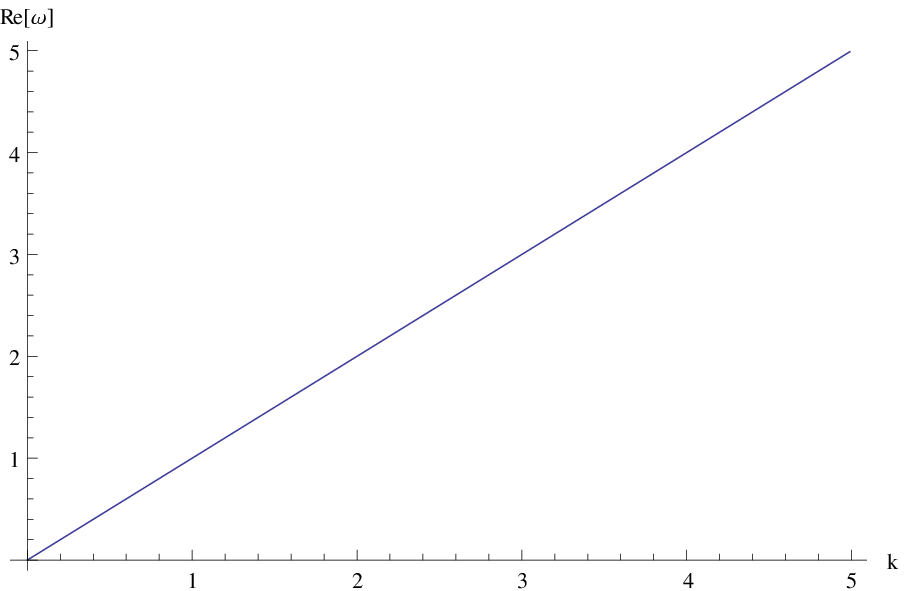}
\caption{\label{fig:normalE3E3}
Linear dispersion relation from our numerics: 
The quasinormal mode with lowest energy of the bulk field $E^3$ in the 
normal phase. This corresponds to the lowest-lying pole of the corresponding
Green's function $G^R_{\mathfrak{e e}}$ of the dynamical electric
field $\mathfrak{e}=\omega \mathfrak{a}_x^3+k\mathfrak{a}_t^3$.
}
}

\paragraph{Broken phase at vanishing spatial momentum $k=0$}
Here we are going to demonstrate that there is a non-trivial gapless mode
appearing in the spectrum of our $(1+1)$-dimensional theory.
We work at finite $W$, $\Phi$ and $\omega$, but vanishing momentum $k=0$.
Then we have three coupled fields ${a_t^1},\, a_t^2,\, a_x^3$, but these
can be reduced to one gauge-invariant field $\hat{a}_x^3$ being dual 
to $\hat{\mathfrak{a}}_x^3$, and obeying a single 
boundary condition (apart from the infalling condition), as discussed in 
section~\ref{sec:methods} above.

We consider the Green's function of the dynamical gauge field
combination $\hat{\mathfrak{a}}_x^3$ in terms of the gauge-invariant
electric field $\hat{\mathfrak{e}}_x^3 = \omega \hat{\mathfrak{a}}_x^3$~\cite{Jensen:2010em}
\begin{equation}
G^R_{\hat{\mathfrak{e}}_x^3\hat{\mathfrak{e}}_x^3} 
=\omega^2 G^R_{\hat{\mathfrak{a}}_x^3\hat{\mathfrak{a}}_x^3} 
= \omega^2 \frac{(\text{vev})}{(\text{source}) }
= \omega^2 \frac{\langle \hat{\mathfrak{a}}_x^3\rangle}{\hat{\mathfrak{j}}_x^3}
=  \omega^2 \frac{{\hat{a}_x^3}-\log (z) z \partial_z {\hat{a}_x^3}}{z \partial_z {\hat{a}_x^3}} \, .
\end{equation}

Due to the presence of a Goldstone mode from the spontaneous
breaking of the $U(1)_3$, we expect that at $T=T_c$ there has to be a gapless mode
at $\omega(k=0)=0$. We follow this mode
to lower temperatures $T<T_c$ keeping the momentum fixed at $k=0$.
The result is shown in figure \ref{fig:zeroKSecondSound}. Indeed the mode
appears at $\omega=0$ and moves down the imaginary frequency axis
as the temperature is lowered. So the Goldstone mode becomes a purely
dissipative mode at lower temperatures, resembling a diffusive mode.\footnote{
A similar mode was observed in the holographic s-wave superconductor
in three spacetime dimensions~\cite{Amado:2009ts}.}
However, it then turns around and
approaches $\omega =0$ again for $T\to0$. Figure \ref{fig:ImwZeroKSecondSound}
shows the temperature-dependence of the imaginary gap in this mode. Note that
this pole seems to either vanish or to become very steep at low temperatures.
In the $T=0.44 T_c$ picture of figure \ref{fig:zeroKSecondSound} it is barely visible.

It may seem surprising that this mode does not acquire a mass gap, i.e.
a gap in real values of $\omega$. This is what one naively expects from
spontaneous symmetry-breaking in presence of a dynamical gauge field:
the Goldstone mode should be eaten by the gauge field, making the latter
massive. But recall that gauge theories in two spacetime dimensions
are peculiar as described in section~\ref{sec:schwinger}. Field theory
tells us that any non-zero
mass for the gauge field would also contribute to the chiral anomaly through the
relation given diagrammatically in figure~\ref{fig:PiIsAnomaly}.
However, holographically the chiral anomaly is dual
to a non-vanishing Chern-Simons term of appropriate dimensionality
in the bulk action~\cite{Erdmenger:2008rm,Banerjee:2008th,Son:2009tf}.
Assuming that there is no other way in which the bulk physics can generate
an anomaly, we conclude that our bulk action~\eqref{eq:action} can not
give rise to massive gauge fields. This assumption seems safe since the chiral
anomaly is a non-perturbative topological quantity in field theory.
It necessarily needs to be dual to a topological quantity in the dual geometry.
A pure Yang-Mills term as in~\eqref{eq:action} can not give rise to topologically
distinct solutions.

\FIGURE{
\label{fig:zeroKSecondSound}
\includegraphics{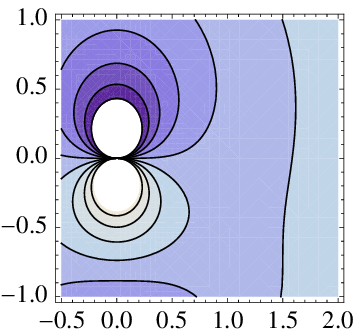}
\includegraphics{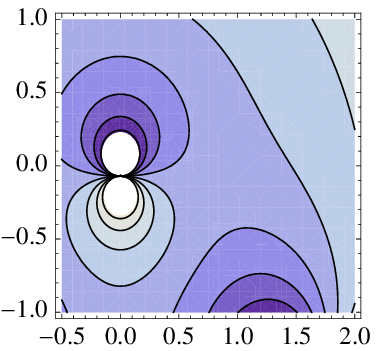}
\includegraphics{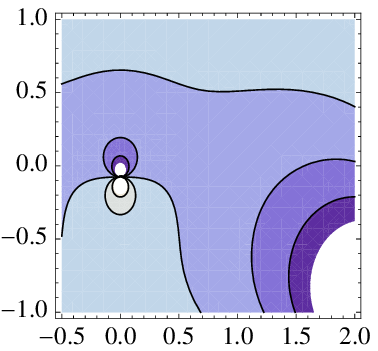}
\includegraphics{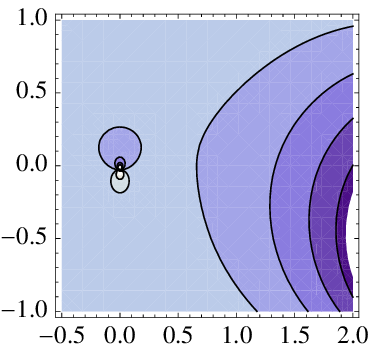}
\includegraphics{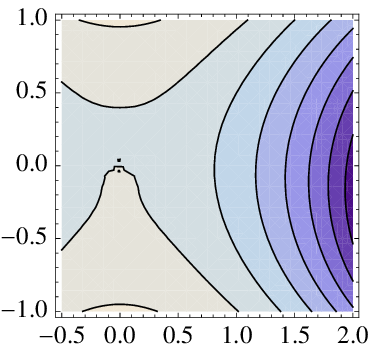}
\caption{Dynamics in the broken phase:
Poles of the correlator $G^R_{\hat{\mathfrak{e}}_x^3\hat{\mathfrak{e}}_x^3}$ (derived from the bulk fluctuation $\hat{a}_x^3$) in the
complex frequency plane are visible in these contour plots of the spectral
function $\propto\text{Im} G^R_{\hat{\mathfrak{e}}_x^3\hat{\mathfrak{e}}_x^3}$. A gapless mode with non-trivial dynamics appears in the broken phase.
The temperature is lowered from $T=T_c$ to $T=0.91 T_c,\, 0.73 T_c,\, 0.57 T_c,\,
0.44 T_c$.
}
}
\FIGURE{
\psfrag{ToverTc}{$\frac{T}{T_c}$}
\psfrag{Imw}{$\mathrm{Im}\omega$}
\includegraphics{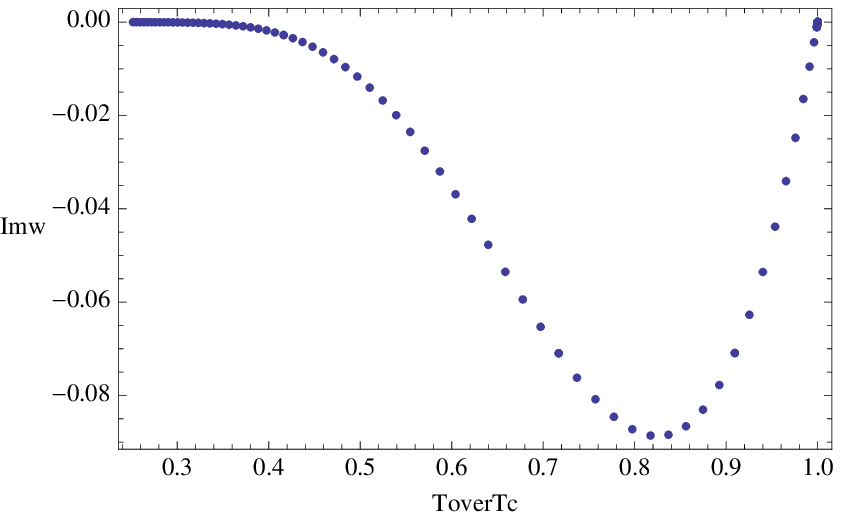}
%\label{fig:imgap}
\label{fig:ImwZeroKSecondSound}
\caption{
The imaginary gap: imaginary part of the frequency of the mode which was gapless
at $T={T_c}$. We keep $k=0$ fixed.}
}
Figures~\ref{fig:imGreens} and~\ref{fig:reGreens} show the two-point function 
of the dynamical field strength $\hat{\mathfrak{e}}^3_{x}$, or in other words the electric field on the boundary.
We find an energy gap of size $\omega_\text{gap}$ in the spectral function, which is presumably
related to the energy needed for breaking a charge carrier out of the
condensate. 

In fact we could employ the picture of a photon propagating 
through a superconductor here. A zero value of that propagator in the 
frequency-gap means that the photons do not propagate. Thus we
indeed see that the $U(1)$-interactions become short-ranged here
although the photons do not become massive.

\FIGURE[h]{
\psfrag{omega}{$\omega$}
\psfrag{exex}{$\text{Im}\, G^R_{\hat{\mathfrak{e}}^3_x\hat{\mathfrak{e}}^3_x}$}
\includegraphics{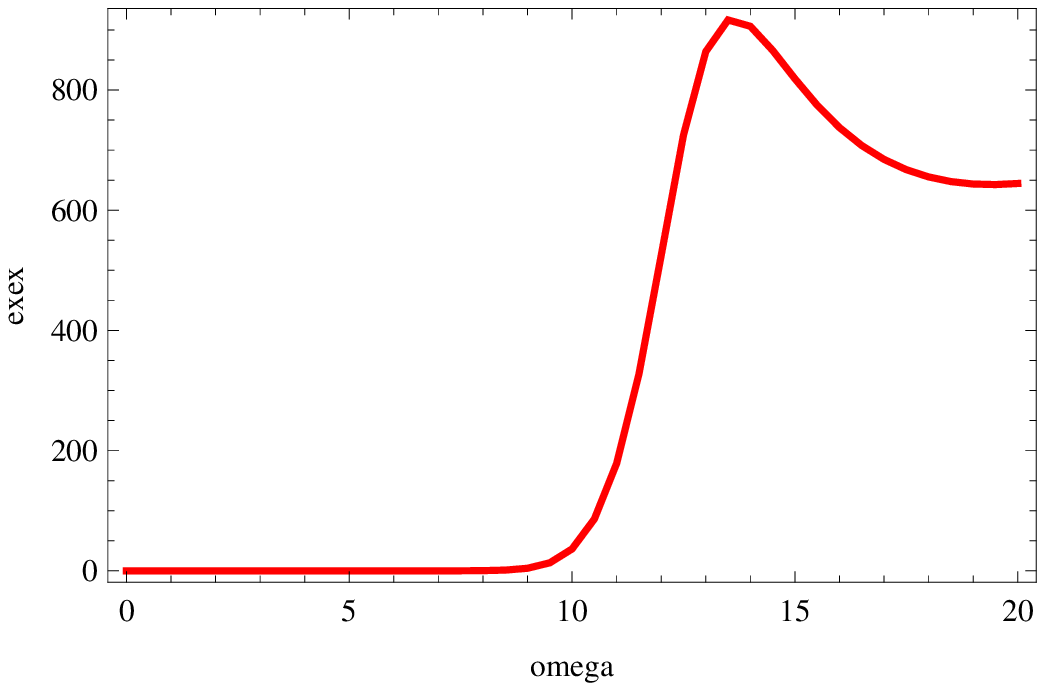}
\caption{\label{fig:imGreens}
Spectral function in broken phase shows frequency-gap $\omega_\text{gap}\approx 10$: Imaginary part of the Green's function of the dynamical electric
field $\hat{\mathfrak{e}}^3_x$ on the boundary as a function of frequency $\omega$
at $T=0.169 T_c$. In the gap the gauge field (our analog of a photon) 
essentially does not propagate. This can be interpreted as the gauge-interaction
(our analog of electromagnetism) becoming short-ranged.}
}
\FIGURE[h]{
\psfrag{omega}{$\omega$}
\psfrag{exex}{$\text{Re}\, G^R_{\hat{\mathfrak{e}}^3_x\hat{\mathfrak{e}}^3_x}$}
\includegraphics{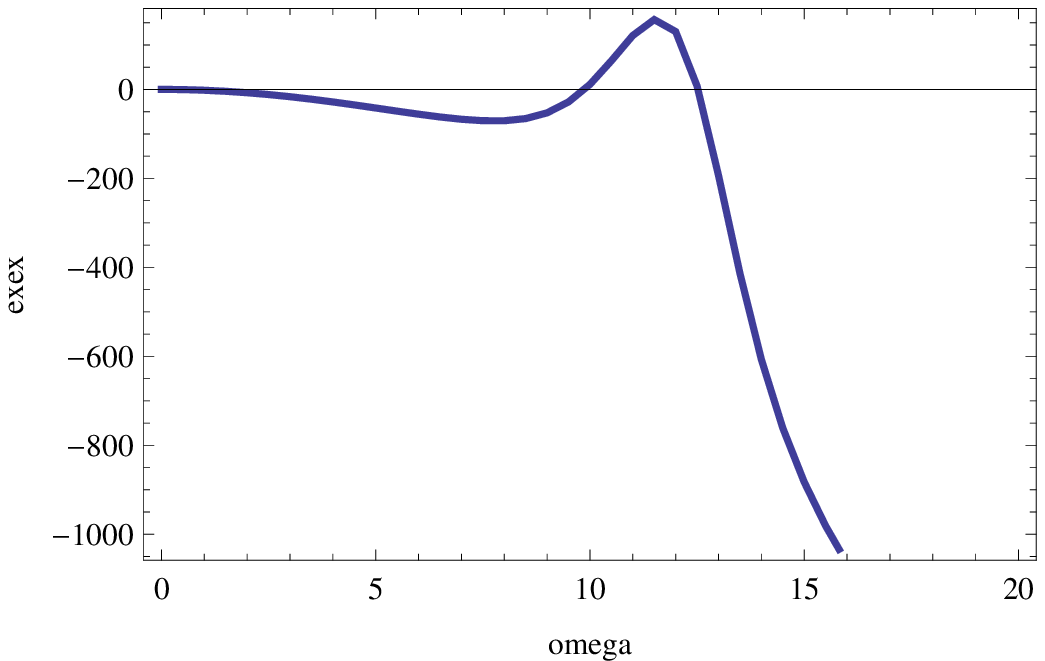}
\caption{\label{fig:reGreens}
Broken phase: Real part of the Green's function of the dynamical electric field
$\hat{\mathfrak{e}}^3_x$ at $T=0.169 T_c$.
}
}

Our currents are not dynamical. However, as suggested in~\cite{Jensen:2010em} we may formally define a conductivity via Ohm's law as 
\begin{equation}\label{eq:formalConductivity}
\sigma = i\frac{\hat{\mathfrak{j}}^3_x}{\omega\langle \hat{\mathfrak{a}}^3_x\rangle} 
% =-i\omega \left({G^R_{\hat{\mathfrak{a}}_x \hat{\mathfrak{a}}_x}}\right)^{-1}
 \, .
\end{equation}
Figure~\ref{fig:formalConductivity} shows this quantity.
There is a prominent resonance-like spike near $\omega=10$ 
in our conventions. From the numerics it is not clear if this spike has finite height.
Note the surprising match between the quantity
$\omega^{-3}G^R_{\hat{\mathfrak{e}}_x \hat{\mathfrak{e}}_x}$
and the Drude model shown in figure~\ref{fig:Drude}, and discussed in appendix~\ref{sec:check}.
We have also tried fitting the formal conductivity given in 
figure~\ref{fig:formalConductivity} using the Drude model
but did not succeed.
\FIGURE{
\includegraphics{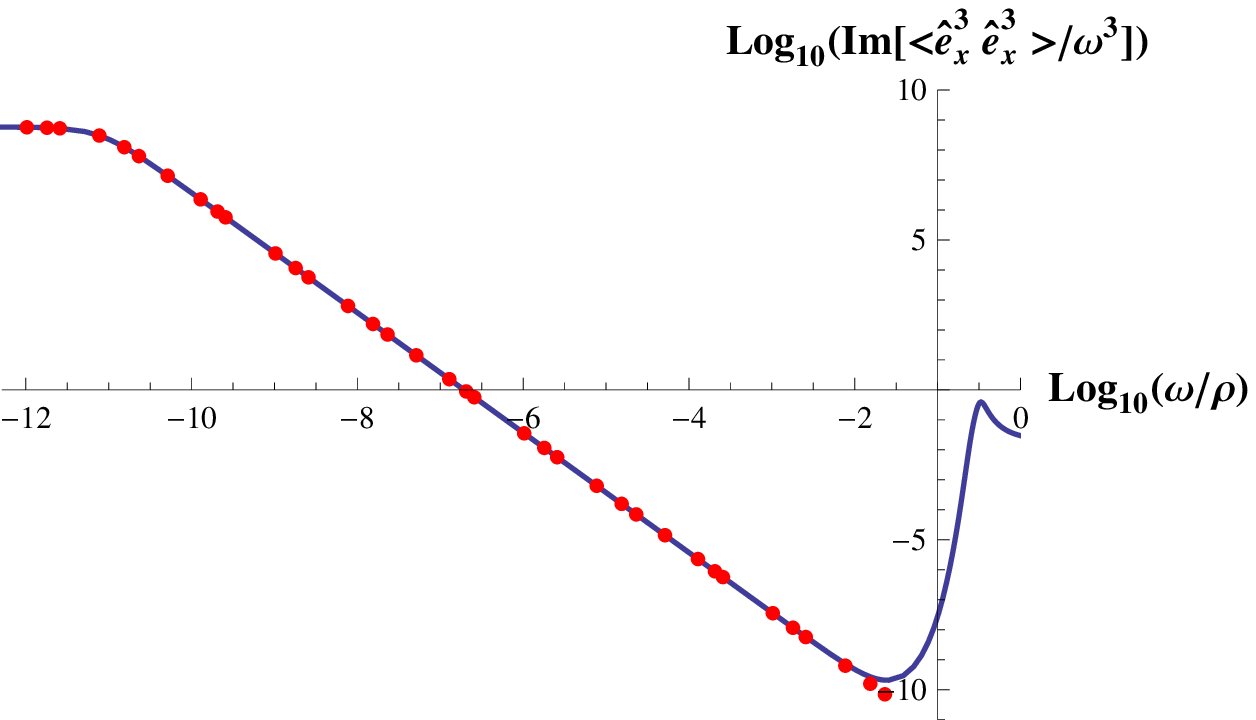}
\caption{ \label{fig:Drude}
Broken phase: Fit of our field strength correlator divided by $\omega^3$ 
to the Drude model. The relevant parameters are
$\sigma_0\approx 5.8288\times 10^8$, $\tau\approx 3.2026\times
10^9$ when $T/T_c\approx0.169$.
}
}
\FIGURE{
\includegraphics{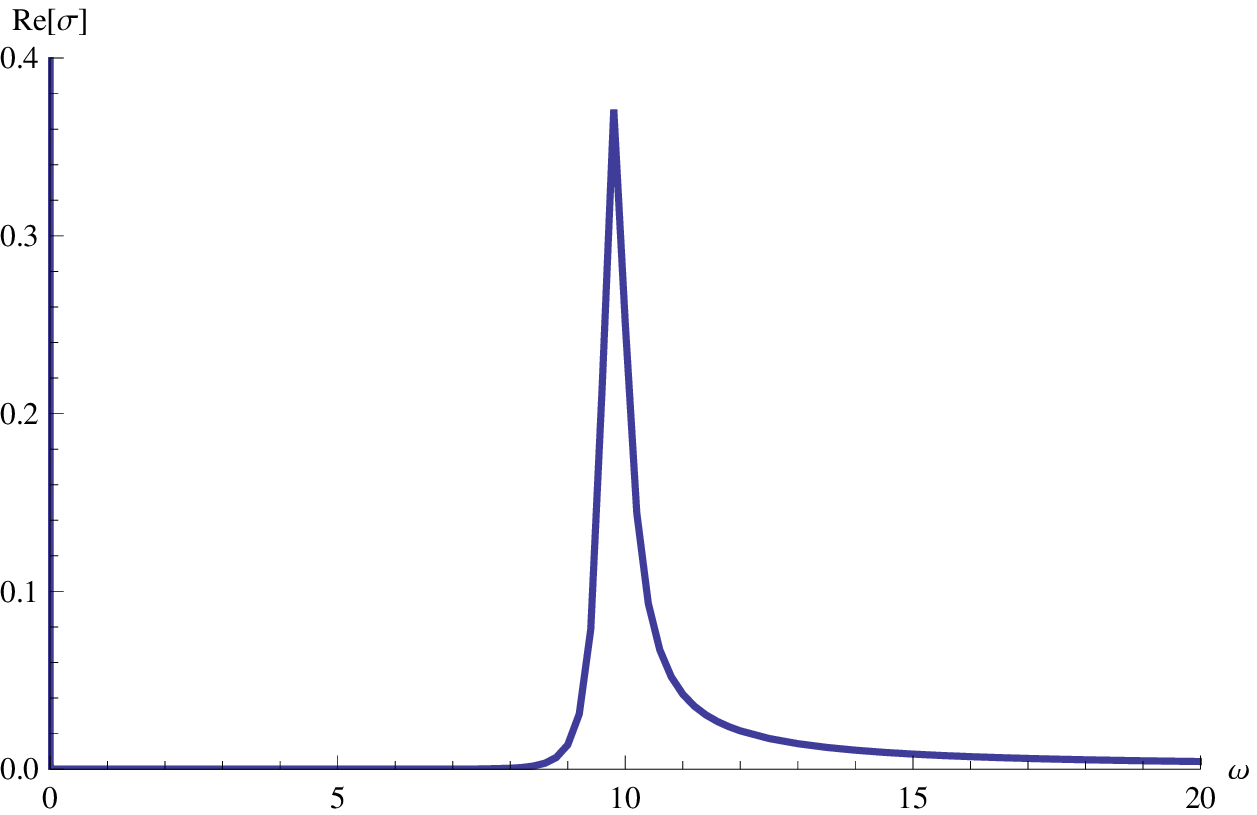}
\caption{ \label{fig:formalConductivity}
Broken phase: The formally-defined conductivity versus frequency 
at $T=0.169T_c$.
}
}
\FIGURE{
\includegraphics{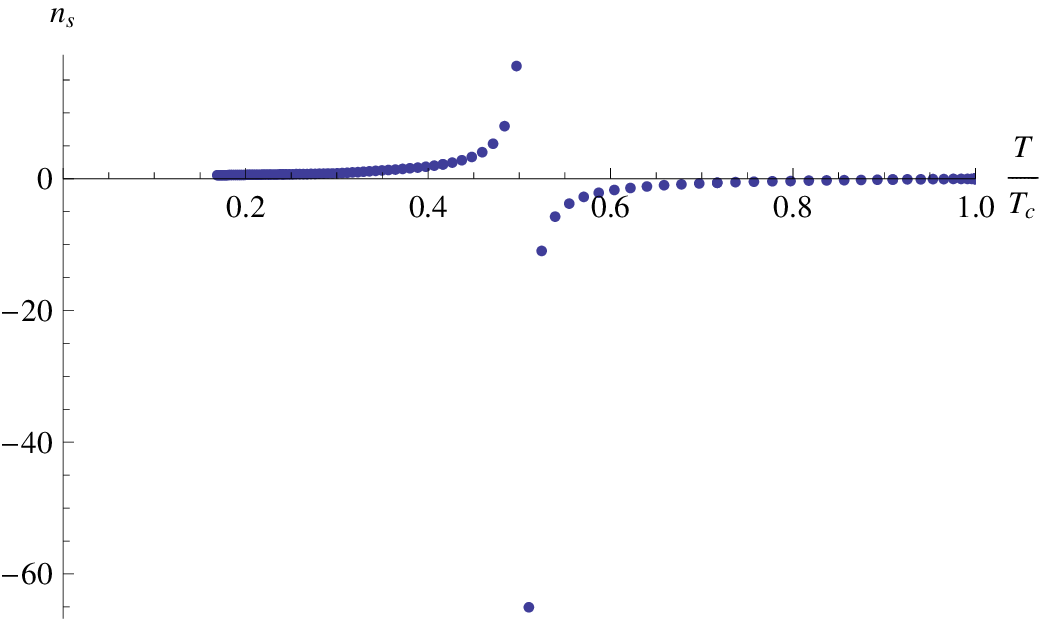}
\caption{\label{fig:alternativeSCDensity}
The formally-defined superconducting density diverges at $T^*\approx 0.5 T_c$
when the alternative interpretation is employed. 
}
}
\FIGURE{
\includegraphics{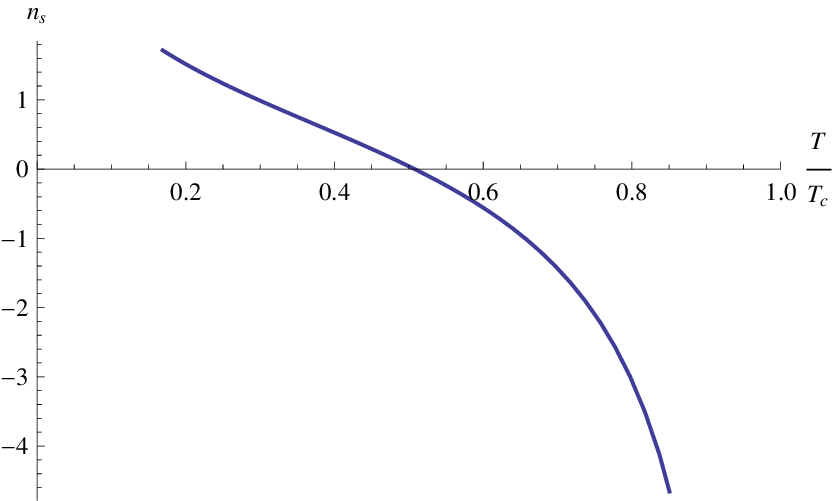}
\caption{\label{fig:SCDensity}
The supercounducting density diverges at $T=T_c$
when the ordinary interpretation is employed. 
It vanishes at a particular $T^*\approx 0.5 T_c$.
}
}
Note that our present setup and field configuration are formally similar to the case
of the holographic s-wave superconductor studied in~\cite{Ren:2010ha}.
But note also the distinct interpretation: in~\cite{Ren:2010ha} the fluctuations
of gauge fields have been interpreted using the standard ordinary quantization.
In particular this interpretation yielded a superconducting charge density
which was negative and divergent near $T_c$. We find exactly the same
divergent behavior of the corresponding superconducting density from our
fluctuations if we use that standard interpretation, see figure~\ref{fig:SCDensity}.
If we choose the alternative quantization interpretation,
the formally-defined superconducting density derived from~\eqref{eq:formalConductivity}
via 
$n_s = \lim\limits_{\omega\to 0}\text{Im}(\omega\sigma)$
seems well-behaved at all frequencies, except for one divergence
at $T^*\approx0.5 T_c$, see figure~\ref{fig:alternativeSCDensity}.
This divergence occurs where the graph in figure~\ref{fig:SCDensity} crossed the real frequency axis. Note that the result displayed in figure~\ref{fig:alternativeSCDensity} is essentially the 
inverse of figure~\ref{fig:SCDensity}.

This somewhat odd behavior of the formally-defined conductivity 
suggests that either we should not interpret $\sigma$
as a conductivity, or our $(1+1)$-dimensional system shows
very peculiar behavior, such as a 
divergent superconducting density at a particular temperature 
$T^*=0.5 T_c$.

\paragraph{Broken phase at non-vanishing spatial momentum $k\neq0$}
It is straightforward to find numerical solutions to the full coupled system of
equations given in \eqref{eq:eoms}. However, the interpretation of these
solutions is not obvious. All the modes are coupled, which means that
in the dual field theory all the operators mix along the renormalization
group flow.

In principle we would like to find quasinormal modes and their dispersion
relations $\omega_{n,\text{QNM}} (k)$ in this system. However, we
can not simply use the prescription given in~\cite{Kaminski:2009dh}.
This prescription needs to be adjusted to the fact that the sources
appear in the logarithmic terms of the bulk field expansions near the
AdS-boundary. We leave this point for future work.

%%%%%%%%%%%%%%%%%%%%%%%%%%%%%%%%%%%%%%%
\section{Discussion} \label{sec:discussion}

In this work we have shown that a black hole in Yang-Mills theory
in AdS$_3$ can grow vector hair, as seen in figure \ref{fig:condensate}.
This solution with a non-trivial profile for the
charged gauge field is thermodynamically preferred at low temperatures,
as seen from the lower free energy in figure \ref{fig:freeenergy}.
This solution corresponds to a boundary theory which develops a
non-zero vacuum expectation value for a dynamical gauge field
$\langle\mathfrak{A}_x\rangle\not=0$. Therefore the local gauge symmetry
of this boundary theory is broken spontaneously, leading us
to interpret this phase as a p-wave superconductor in $(1+1)$
dimensions. Formally the symmetry breaking in presence of our dynamical gauge field 
occurs just like in the setups with non-dynamical gauge field which yield superfluids. 
However, we are not able to directly compare the two processes within our gravity model since we
are restricted to the alternative quantization, i.e. to the dynamical gauge field case. See
section \ref{sec:altBC} for an extended discussion of this point.

Further we have studied the dynamics of this model and have
shown that at least one non-trivial gapless mode exists in the broken phase.
This may allow a low-energy effective analysis of this setup in what we defined in section~\ref{sec:altNormalHydro} as "non-equilibrium field dynamics" (this is analogous
to the hydrodynamic description in the standard quantization).
The origin of this gapless mode is the Goldstone
boson which appears at the transition temperature. Naively
we would expect this mode to give mass to our $U(1)_3$ gauge field,
just like the photon becomes massive in a traditional superconductor.
However, our numerical studies
suggest that our gauge field does not acquire a mass, in contrast to 
our naive expectation. Instead, our Goldstone mode below the transition
acquires a larger and larger damping
at lower temperatures, as was seen in figure \ref{fig:ImwZeroKSecondSound}.

This fact again confirms our interpretation of these two-point functions
as correlators of dynamical gauge fields (i.e. gauge field
propagators) in the following way:
We know from the Schwinger model that in two spacetime
dimensions a $U(1)$ gauge field can only have mass if the theory
also has a chiral anomaly, see discussion in section \ref{sec:schwinger}.
However, we have forced our anomaly
to vanish by forcing the Chern-Simons coupling to vanish in the action~\eqref{eq:action}.
Therefore we conclude that our gauge bosons can not acquire a
mass in the present setup. See section~\ref{sec:numerics} 
for a detailed discussion. It would be very interesting to
investigate the action \eqref{eq:action} after addition of a Chern-Simons term.
We suspect that this will allow a mass generation by a Higgs-like mechanism
below the critical temperature.\footnote{In the conclusions of~\cite{Jensen:2010em} it was noted that for such a theory a perturbative instability is ruled out due to flatness of the gauge connection. However, non-perturbatively there may exist a thermodynamically preferred phase with a (vector) condensate.}
In fact, the size of the condensate
may be linked to the size of the anomaly, see~\cite{Zayas:2011dw}
for a similar setup in AdS$_5$. 
The significance of the 
formally-defined~\cite{Jensen:2010em} 
conductivity~\eqref{eq:formalConductivity} requires more investigation.

We also see the gauge boson propagator (virtually) vanish in the superconducting
phase at low frequencies, see figure~\ref{fig:imGreens}. This suggests
that the gauge interactions mediated by our field theory gauge boson $\hat{\mathfrak{a}}_x^3$ (our analog of the photon)
become short-ranged. In other words, we see that our analog of the 
electromagnetic interaction becomes short-ranged. In this context
it would be interesting to perform a calculation at finite momentum,
which may allow to derive the penetration depth of this superconductor.
As a general conclusion this holographic superconductor in the alternative formulation
shows signatures which are strikingly similar to the ones that are known in the ordinary formulation.
For example the gap in our gauge boson propagator (figure~\ref{fig:imGreens}) resembles
the (conductivity) gap observed in the current-current two-point functions in the ordinary formulation, see e.g.~\cite{Hartnoll:2008vx}.
However, we stress again, that the interpretation of these signatures and quantities is different in
the ordinary and the alternative formulations. For example, the gap in the ordinary current-current correlators
is interpreted as a gap in the (single electron) conductivity. In our alternative setup the gap in the corresponding
gauge-field two-point function is interpreted differently as explained at the beginning of this paragraph.

Although our action~\eqref{eq:action} preserves parity, our ground state
solution spontaneously breaks this symmetry. With this property our system 
may serve as a toy model for unconventional superconductors, which violate parity
in their ground state through the order parameter. An experimentally 
well-studied example for such a system in $2+1$ dimensions is the strontium ruthenate compound $Sr_2 RuO_4$~\cite{Maeno:1994na}. In this context it may also 
be interesting to consider an analog to our setup in AdS$_4$.
Note that in the ordinary quantization the obvious (p+ip)-wave
solution which breaks parity in AdS$_4$
is thermodynamically not favored according to the analysis in~\cite{Zayas:2011dw}.

None of our fluctuations indicates an instability of the p-wave phase. However,
a study of the full mode spectrum of the coupled system at finite
spatial momentum turns out to be involved.
In particular this requires an extension of the (determinant)
method proposed in~\cite{Kaminski:2009dh} in order to account for the logarithmic terms
in the AdS$_3$-asymptotics. This is postponed to future work.

Our paper leaves some other loose ends which are worthwhile being picked up.
In particular one could repeat our analysis in AdS$_4$,
where both quantizations are possible, and thus
can be compared directly. This should allow to make the relation of
Green's functions in the alternative quantization to the ones in the
ordinary quantization explicit.
In AdS$_4$ we also have a magnetic field, which can destroy the ordered state,
and a dynamical response should be observable. So the Meissner-effect
could be observed directly. 
For this purpose the
authors of~\cite{Domenech:2010nf,Montull:2012fy,Silva:2011zz,Montull:2011im}
provide stationary solutions (in the s-wave superconductor), while we suggest
to examine the dynamics of their backgrounds via the bulk field fluctuations along the
lines of our present paper.

In analogy to the standard hydrodynamic study of higher-dimensional setups,
see for example~\cite{Policastro:2002se,Bhattacharyya:2008jc},
one could study the alternatively quantized setups at low frequency and momentum. Such a study should yield results resembling those of traditional thermal quantum
field theory, and field dynamics out of equilibrium. Due to the simplicity of
our setup one may hope for analytic results in the spirit of~\cite{Herzog:2009ci}.

In fact our investigations suggest that the alternative interpretation may
be interesting for a wide class of holographic fluids: 
It should be possible to study each of the holographic
fluids which have been investigated in AdS$_4$, but now with the Neumann boundary condition, i.e. in the alternative quantization. Our approach should be applicable to all holographic fluids, such as holographic quark-gluon plasma 
and all superfluids, and superconductors (s-wave, p-wave, d-wave),
which have been found in AdS$_4$.

%%%%%%%%%%%%%%%%%%%%%%%% A C K N O W L E D G M E N T S
\section*{Acknowledgements}
It is a pleasure to thank M.~Ammon, T.~Faulkner, S.S.~Gubser, S.~Hartnoll,
C.P.~Herzog, A.~Karch, K.~Landsteiner, J.~Ren, M.~Roberts, J.~Scholtz, D.T.~Son, A.~Starinets, H.~Verlinde,
and especially K.~Jensen for valuable discussions. We thank the referee for helpful comments.
HQZ is grateful for Prof. Rong-Gen Cai's encouragement. 
XG is supported by the MPG-CAS Joint Doctoral Promotion Programme.
MK is currently supported by the US Department of Energy under contract number DE-FGO2-96ER40956, and in part by the National Science Foundation under Grant No. NSF PHY11-25915. 
HQZ was supported in part by the National Natural Science Foundation of China (No.10821504,
No.10975168 and No.11035008), and in part by the Ministry of Science and
Technology of China under Grant No. 2010CB833004. 
HBZ is supported by the Fundamental Research Funds for the Central Universities (Grant No.1107020117) and
the China Postdoctoral Science Foundation (Grant No. 20100481120).
MK thanks the KITP, Santa Barbara for kind
hospitality during part of this project.
%%%%%%%%%%%%%%%%%%%%%%%% A P P E N D I X
\begin{appendix}
%______________________________________________
\section{Gauge-Invariant and Gauge-Covariant Fields}
\label{sec:covariantFields}

\subsection{Broken Phase}
In the broken phase (superconducting phase), the gauge
transformation of the perturbations are given by
 \be\label{gauge} \dl(e^{-i\om t+ikx}a^a_\mu)=\pa_\mu(e^{-i\om
 t+ikx}\al^a)+e^{-i\om t+ikx}\eps^{abc}A^b_\mu \al^c.\ee
 where, $\al^a$ can not depend on $z$ because we have chosen an
 axial gauge. Therefore $\al^a$ has
 constant components.

One way of writing these gauge transformations of $a^a_\mu$ from
\eqref{gauge} explicitly is
\be\label{matrix} \dl\left(
                 \begin{array}{c}
                   { a^1_x} \\
                   a^2_x \\
                   a^3_x \\
                   { a^1_t} \\
                   { a^2_t} \\
                   a^3_t
                 \end{array}
               \right)=
               \left(
                 \begin{array}{ccc}
                   { ik} & { 0} & { 0} \\
                   0 & ik & -W(z) \\
                   0 & W(z) & ik \\
                   {-i\om} & {-\Phi(z)} & {0} \\
                   {\Phi(z)} & {-i\om} & {0}\\
                   0 & 0 & -i\om
                 \end{array}
               \right)\times\left(
                 \begin{array}{c}
                   \al^1 \\
                   \al^2 \\
                   \al^3
                 \end{array} 
               \right)\, .\ee 
Actually, there are various gauge invariant linear combinations for
one typical perturbation. For example, we can calculate the linear
gauge invariant combinations of $a^3_x$ in general form as follows:
 \be \hat{a}^3_x=a^3_x+h\, a^1_x+j \,a^2_x+m \,a^1_t+n\, a^2_t+p \,a^3_t.\ee
where $(h, j, m, n, p)$ are coefficients without $a^a_\mu$. Because
of $\dl \hat{a}^3_x=0$, we can have 7 types of gauge invariant
combinations:
 \begin{enumerate}
 \item $ h=j=0,\quad p= \frac{k}{\omega },m= \frac{W(z) \Phi
   (z)}{\Phi (z)^2-\omega ^2},n= -\frac{i \omega
   W(z)}{\omega ^2-\Phi (z)^2} $;
 \item $h=m=n=0,\quad p= -\frac{W(z)^2-k^2}{k \omega },j=
   \frac{i W(z)}{k}$;
 \item $h=p=0, \quad m= -\frac{\left(W(z)^2-k^2\right) \Phi
   (z)}{W(z) \left(\omega ^2-\Phi (z)^2\right)},n=
   \frac{i \left(k^2 \omega -\omega
   W(z)^2\right)}{W(z) \left(\omega ^2-\Phi
   (z)^2\right)},j= \frac{i k}{W(z)}$;
 \item $j=m=0,\quad h= \frac{W(z) \Phi (z)}{k \omega },p=
   \frac{k}{\omega },n= -\frac{i W(z)}{\omega
   }$;
 \item $j=n=0,\quad h= \frac{\omega  W(z)}{k \Phi (z)},p=
   \frac{k}{\omega },m= \frac{W(z)}{\Phi
   (z)}$;
 \item $m=p=0,\quad h= -\frac{\left(k^2-W(z)^2\right) \Phi
   (z)}{k \omega  W(z)},n= \frac{i
   \left(k^2-W(z)^2\right)}{\omega  W(z)},j=
   \frac{i k}{W(z)} $;
 \item $n=p=0,\quad h= -\frac{\omega  \left(k^2-W(z)^2\right)}{k
   W(z) \Phi (z)},m= -\frac{k^2-W(z)^2}{W(z) \Phi
   (z)},j= \frac{i k}{W(z)}$.
 \end{enumerate}
In fact, one can also construct the linear gauge-invariant
combinations for other perturbations in the same spirit. We will not
show them explicitly here. 

\subsection{Normal Phase}
In the normal phase, only part of the $SU(2)$ is broken and a  $U(1)_3$ remains intact. 
Then the fluctuations $a_{\mu}^a$ transform as
\begin{equation}
\delta a_{\mu}^a = \partial_{\mu} \lambda^a + \epsilon^{abc}(A_{\mu}^b \lambda^c + a_{\mu}^b\Lambda^3 \delta_3^c) \, ,
\end{equation}
where $\Lambda^3$ is the gauge transformation parameter of the $U(1)_3$. 
From this we obtain
\begin{eqnarray}
\label{transform}
\delta a_t^1 &=& -i \omega \lambda^1 - \Phi(z) \lambda^2 + a_t^2 \Lambda^3\, ,\nonumber\\
\delta a_t^2 &=& -i \omega \lambda^2 + \Phi(z) \lambda^1 - a_t^1 \Lambda^3\, ,\nonumber\\
\delta a_t^3 &=& -i \omega \lambda^3\, ,\nonumber\\
\delta a_x^1 &=& i k \lambda^1 + a_x^2 \Lambda^3\, ,\nonumber\\
\delta a_x^2 &=& i k \lambda^2 - a_x^1 \Lambda^3\, ,\nonumber\\
\delta a_x^3 &=& i k \lambda^3\, .
\end{eqnarray}
Define $e_t^{\pm} =a_t^1 \pm  i a_t^2 $, $e_x^{\pm} =a_x^1 \pm i a_x^2$ which are in the fundamental representation of $U(1)_3$, and $a_{\mu}^3$ is in the adjoint. 
Using the above transformation,  we obtain
\begin{eqnarray}
\delta e_t^+ &=& \partial_{t} \lambda^{+} + i \Phi (x) \lambda ^{+} - i \Lambda ^3 e_t^{+}\, , \nonumber\\
\delta e_t^- &=& \partial_{t} \lambda^{-} - i \Phi (x) \lambda ^{-} + i \Lambda ^3 e_t^{-}\, , \nonumber\\
\delta e_x^+ &=& \partial_x \lambda^{+} - i \Lambda ^3 e_x^{+}\, ,\nonumber \\
\delta e_x^- &=& \partial_x \lambda^{-} + i \Lambda ^3 e_x^{-}\, , \nonumber\\
\delta a_{\mu}^3 &=& \partial _{\mu} \lambda^3\, .
\end{eqnarray}
From the $a_t^3$ and $a_x^3$, we can construct a $U(1)_3$-invariant combination with the help of Fourier transformation ($e^{-i \omega t + i k x} f(t,x) \rightarrow f(\omega, k)$ as before), that is
\begin{equation}
e_3 = k a_t^3 + \omega a_x^3 \, ,
\end{equation}
which leads to $\delta e_3 = -i k \omega \lambda^3 + i k \omega \lambda^3 = 0$.

On the other hand, we can build $U(1)_3$ gauge-covariant combinations 
for $e_t^{\pm}$ and $e_x^{\pm}$,
\begin{eqnarray}
e_L^{+} &=& (\Phi(z) - \omega) e_{x}^{+} -  k e_{t}^{+}, \\
e_L^{-} &=& (\Phi(z) + \omega) e_{x}^{-} +  k e_{t}^{-}.
\end{eqnarray}
This leads to
\begin{equation}
\delta e_L^{+} = (\Phi(z) - \omega) \delta e_{x}^{+} - k \delta e_{t}^{+} 
=- i \Lambda^3 e_L^+ \, ,
\end{equation}
and
\begin{equation}
\delta e_L^- = (\Phi(z) + \omega) \delta e_{x}^{-} + k \delta e_{t}^{-} 
= i \Lambda^3 e_L^- \, .
\end{equation}
With these gauge-covariant fields $e_L^+$ and $e_L^-$ and the gauge-invariant field $e_3$ in the normal phase, we can rewrite the on-shell action. From the on-shell action we can see that $e_3$ decouples from all other fields while $e_L^+$ and $e_L^-$ are coupled to each other.

%______________________________________________
\section{Ordinary Interpretation}\label{sec:check}

\subsection{Drude Model}
The quantity $\omega^{-3}G^R_{\hat{\mathfrak{e}}_x \hat{\mathfrak{e}}_x}$ 
is identical to the conductivity one would define in 
the ordinary interpretation of our setup. Therefore, 
we tried fitting this using the Drude model
\begin{eqnarray}
\rm Re(\sigma_{Drude})=\frac{\sigma_0}{1+\omega^2\tau^2}.
\end{eqnarray}
where, $\sigma_0=nq^2\tau/m$ is the DC conductivity of
$\sigma_{xx}$, while $n$, $q$, $m$, $\tau$ are respectively the
electron's number density, charge, mass and the mean free time
between collisions.
As seen from figure~\ref{fig:Drude} this works surprisingly well at
low frequencies. We have no good explanation for this agreement.

\subsection{Superconducting Density}
Here we assume that the ordinary interpretation of sources as 
fixed boundary gauge fields is correct. In this case our two-point functions
would be correlators of dynamical currents. 
In particular this would allow us to compute the superconducting density
using the formula
\begin{equation}
n_s =\lim\limits_{\omega\to0}\text{Re}\, G_{\mathfrak{j j}}^R(\omega; k=0) \, .
\end{equation}
Here the Green's functions is obtained from the infalling solution of
the bulk field $\hat{a}^3_x$ given in~\eqref{eq:aHat}.
The result is shown in figure~\ref{fig:SCDensity}. We observe the same 
negative divergent behavior as seen in~\cite{Ren:2010ha}. We take this
as a numerical evidence for the fact that the alternative interpretation 
should be applied to this kind of setup. See section~\ref{sec:fluctuations}
for a detailed discussion of the alternative interpretation.

\end{appendix}

%%%%%%%%%%%%%%%%%%%%%%%% R E F E R E N C E S
\bibliography{ads3.v3}{}
\bibliographystyle{JHEP}     % bibtex style file JHEP

\end{document}